\documentclass[reprint,amsmath,amssymb,aps,prx,superscriptaddress]{revtex4-2}

\usepackage[version=3]{mhchem}
\usepackage{amsfonts}
\usepackage{graphicx}
\usepackage{dcolumn}
\usepackage{bm}
\usepackage[table]{xcolor}
\usepackage{booktabs}
\usepackage[colorlinks=true,linkcolor=blue,citecolor=blue,urlcolor=black]{hyperref}

\newcommand{\bx}{\mathbf{x}}
\newcommand{\pp}{p_\phi}
\newcommand{\pt}{p_\theta}

\begin{document}

\title{Scaling Autoregressive Models for Lattice Thermodynamics}

\author{Xiaochen Du}
\affiliation{Department of Chemical Engineering, Massachusetts Institute of Technology, Cambridge, MA 02139, USA}
\author{Juno Nam}
\affiliation{Department of Materials Science and Engineering, Massachusetts Institute of Technology, Cambridge, MA 02139, USA}
\author{Sulin Liu}
\affiliation{Department of Materials Science and Engineering, Massachusetts Institute of Technology, Cambridge, MA 02139, USA}
\author{Rafael G\'omez-Bombarelli}
\email[Corresponding author: ]{rafagb@mit.edu}
\affiliation{Department of Materials Science and Engineering, Massachusetts Institute of Technology, Cambridge, MA 02139, USA}

\date{\today}

\begin{abstract}
Predicting how materials behave under realistic conditions requires understanding the statistical distribution of atomic configurations on crystal lattices, a problem central to alloy design, catalysis, and the study of phase transitions. Traditional Markov-chain Monte Carlo sampling suffers from slow convergence and critical slowing down near phase transitions, motivating the use of generative models that directly learn the thermodynamic distribution. Existing autoregressive models (ARMs), however, generate configurations in a fixed sequential order and incur high memory and training costs, limiting their applicability to realistic systems. Here, we develop a framework combining any-order ARMs, which generate configurations flexibly by conditioning on any known subset of lattice sites, with marginalization models (MAMs), which approximate the probability of any partial configuration in a single forward pass and substantially reduce memory requirements. This combination enables models trained on smaller lattices to be reused for sampling larger systems, while supporting expressive Transformer architectures with lattice-aware positional encodings at manageable computational cost. We demonstrate that Transformer-based any-order MAMs achieve more accurate free energies than multilayer perceptron-based ARMs on both the two-dimensional Ising model and CuAu alloys, faithfully capturing phase transitions and critical behavior. Overall, our framework scales from $10 \times 10$ to $20 \times 20$ Ising systems and from $2 \times 2 \times 4$ to $4 \times 4 \times 8$ CuAu supercells at reduced computational cost compared to conventional sampling methods.
\end{abstract}

\maketitle

\section{Introduction}

Understanding the equilibrium distribution of states is a prerequisite for predicting realistic materials behavior under experimental conditions. At finite temperatures, material properties are rarely determined by a single structure. Across length scales and material classes, thermodynamic behavior is governed by ensembles of configurational states and their statistical weights. Molecular systems such as proteins and polymers sample conformational ensembles that control thermodynamic and kinetic response~\cite{chandler1987introduction, tuckerman2023statistical, yang_learning_2024, xu_reactive_2024, nam_transferable_2025}, while catalyst surfaces reconstruct and adsorb species as functions of reactant activities, temperature ($T$), and chemical potentials ($\mu$), often spanning nanometer length scales~\cite{seh_combining_2017, bruix_first-principles-based_2019,li_origin_2022,shen_deciphering_2023,du_machine-learning-accelerated_2023,owen_low-index_2024,lee_machine-learning-accelerated_2025,du_accelerating_2025}. These principles likewise determine phase behavior and functional response in disordered alloys~\cite{sheriff_quantifying_2024, sheriff_chemical-motif_2024, sheriff_machine_2025}, complex oxides~\cite{peng_stability_2022,lunger_towards_2024,peng_data-driven_2024, kempen_inverse_2025}, and solid interfaces~\cite{xiao_understanding_2020,spotte-smith_toward_2022,ou_atomistic_2024,ou_non-arrhenius_2025}.

In an important subset of these systems, configurational disorder can be represented by discrete occupational degrees of freedom on a periodic lattice $\bx$. For instance, compositional disorder in substitutional alloys~\cite{han_multifunctional_2024, sheriff_chemical-motif_2024,chun_learning_2024} or configurational degrees of freedom in polymers~\cite{gartner_modeling_2019}. Each lattice site $x_1, x_2, ..., x_L$ in $\bx$ of length $L$ can adopt a discrete state, such as a species identity or spin orientation. In the (semi-)grand canonical ensemble, the energy $E(\bx)$ of a lattice configuration and the occupation number $N_i(\bx)$ of species $i$ define Boltzmann weights $f(\bx) \propto \exp[-\beta(E(\bx) - \sum_i \mu_i N_i(\bx))]$, where $\mu_i$ are chemical potentials, $\beta = \frac{1}{k_{\text{B}}T}$, and $k_{\text{B}}$ is the Boltzmann constant. Sampling from these distributions is commonly performed with Markov-chain Monte Carlo (MCMC) algorithms, paired with energy evaluations from first-principles calculations, cluster expansions, classical potentials, and more recently, machine learning force fields (MLFFs)~\cite{van_de_walle_alloy_2002, takeuchi_new_2017,van_der_ven_first-principles_2018,chang_clease_2019, angqvist_icet_2019,niu_multi-cell_2019, du_machine-learning-accelerated_2023}. While powerful, these sampling approaches are inherently sequential and suffer from critical slowing down near phase transitions. They also become increasingly expensive when phase diagrams must be mapped across many $(T,\mu)$ conditions or when large supercells are required to capture long-range order or to mitigate finite-size artifacts.

Machine learning-based approaches offer complementary strategies to accelerate sampling of lattice Boltzmann distributions, broadly falling into two classes: amortized generative models trained on pre-generated Monte Carlo samples~\cite{tuo_scalable_2025, herron_inferring_2024, lee_exponentially_2025}, and neural samplers trained directly against the Hamiltonian~\cite{holderrieth_leaps_2025, ou_discrete_2025, woo_energy-based_2025, guo_proximal_2025, zhu_mdns_2025}. While these approaches have demonstrated promising performance on canonical models such as the Ising and Potts systems~\cite{holderrieth_leaps_2025, woo_energy-based_2025, guo_proximal_2025, zhu_mdns_2025}, they face practical challenges for the materials chemistry applications motivating this work. Amortized inference models typically use denoising diffusion or flow-matching-based approaches that lack tractable likelihoods in practice, making direct free-energy estimation infeasible. As a result, free energies must be inferred through histogram reweighting or density-of-states estimation applied to generated samples, which can be computationally intensive and scale poorly in high-dimensional $(T, \mu_1, \mu_2, \ldots)$ parameter spaces for phase diagrams of multi-component systems~\cite{tuo_scalable_2025, herron_inferring_2024, lee_exponentially_2025}. Energy-based neural samplers, in contrast, are generally trained for a single thermodynamic condition $(T, \mu)$, limiting their utility for workflows requiring systematic exploration of many thermodynamic state points, as in catalyst design under varying gas-phase chemical potentials or alloy phase boundary determination~\cite{holderrieth_leaps_2025, ou_discrete_2025, woo_energy-based_2025, guo_proximal_2025, zhu_mdns_2025}.

Among these lattice distribution-learning approaches, autoregressive models (ARMs) are particularly attractive because they provide exact normalized likelihoods $p(\bx)$, enabling direct free energy estimation at arbitrary thermodynamic conditions. Fixed-order (FO) ARMs have been applied to lattice Boltzmann distributions~\cite{wu_solving_2019, nicoli_asymptotically_2020, damewood_sampling_2022}, factorizing $p(\bx)$ into a product of conditionals along a predetermined site ordering. However, two limitations hinder their application to material systems at scale: (i) the fixed ordering precludes arbitrary conditional generation, which is important for applications like catalyst design~\cite{chun_active_2024}, and (ii) training requires backpropagation through the full generation sequence, incurring \emph{O}($L^2$) memory cost, where $L$ is the number of lattice sites, limiting accessible lattice sizes and architectural expressiveness.

In this work, we addressed these limitations by combining any-order (AO) autoregressive models with marginalization models (MAMs) tailored for lattice thermodynamics. Inspired by recent developments in AO generative modeling~\cite{liu_generative_2024}, we trained conditional models that, unlike fixed-order ARMs, can generate any site given any subset of known sites, enabling flexible masking (holding a chosen subset of sites fixed) and generation of the remainder. In parallel, we trained neural networks that approximate arbitrary marginals $p(\bx_\mathcal{S})$, that is, the probability of a partial configuration on any subset $\mathcal{S}$ of sites, summed over all unobserved ones. These marginals provide both fast likelihood surrogates and a mechanism to generate training samples via blockwise Gibbs sampling, which iteratively refreshes batches of sites by drawing from their conditional distributions given the rest. The joint training objective enforces consistency between these conditionals and marginals and is compatible with the energy-based training approach used in materials contexts. When direct training on large lattices is impractical, the any-order formulation naturally enables out-painting, a technique from generative image modeling in which generation is extended to new regions by conditioning on adjacent known sites; here we adapted it to transfer models trained on smaller lattices to larger supercells without retraining, analogous to how MLFFs trained on small cells are deployed for large-scale simulations~\cite{owen_low-index_2024}.

We demonstrated our framework on two representative systems: (i) the two-dimensional Ising model, a canonical testbed for critical phenomena, and (ii) the CuAu alloy described by a cluster expansion as a realistic system with multiple stable phases. Across these systems, we showed that:
\begin{enumerate}
    \item Transformer-based MAMs achieve superior performance compared to multilayer perceptron (MLP)-based ARMs, with higher effective sample sizes, more accurate free energies, and better capture of complex phase behavior. For CuAu systems, MAM Transformers successfully capture all three ordered intermetallic phases ($\ce{Cu3Au}$, $\ce{CuAu}$, and $\ce{CuAu3}$), while ARM MLPs systematically miss or incorrectly predict phases, demonstrating the importance of architecture choice for realistic materials systems.
    \item Any-order models enable powerful out-painting strategies that transfer trained models from smaller to larger lattices (e.g., $10 \times 10$ to $15 \times 15$ and $20 \times 20$ Ising, and $4 \times 4 \times 4$ to $4 \times 4 \times 8$ CuAu) without retraining, independent of the specific network architecture. Notably, out-painting from sufficiently large source models (e.g., $15 \times 15$ to $20 \times 20$ Ising) achieves thermodynamic accuracy comparable to or even exceeding directly trained models at zero additional training cost.
    \item Once trained, our models in many cases provide orders-of-magnitude speedups over MCMC, Wang--Landau, and metadynamics sampling, generating samples without requiring per-sample energy evaluations. This offers favorable cost--accuracy trade-offs when many $(T,\mu)$ conditions or larger supercells are required, with the upfront training investment quickly amortized across multiple thermodynamic conditions.
\end{enumerate}
These results establish a path toward scalable, flexible generative models for lattice thermodynamics, opening applications in modeling alloy phase behavior, surface reconstructions, and material interfaces.

\section{Results}
\begin{figure*}
    \centering
    \includegraphics[width=\textwidth]{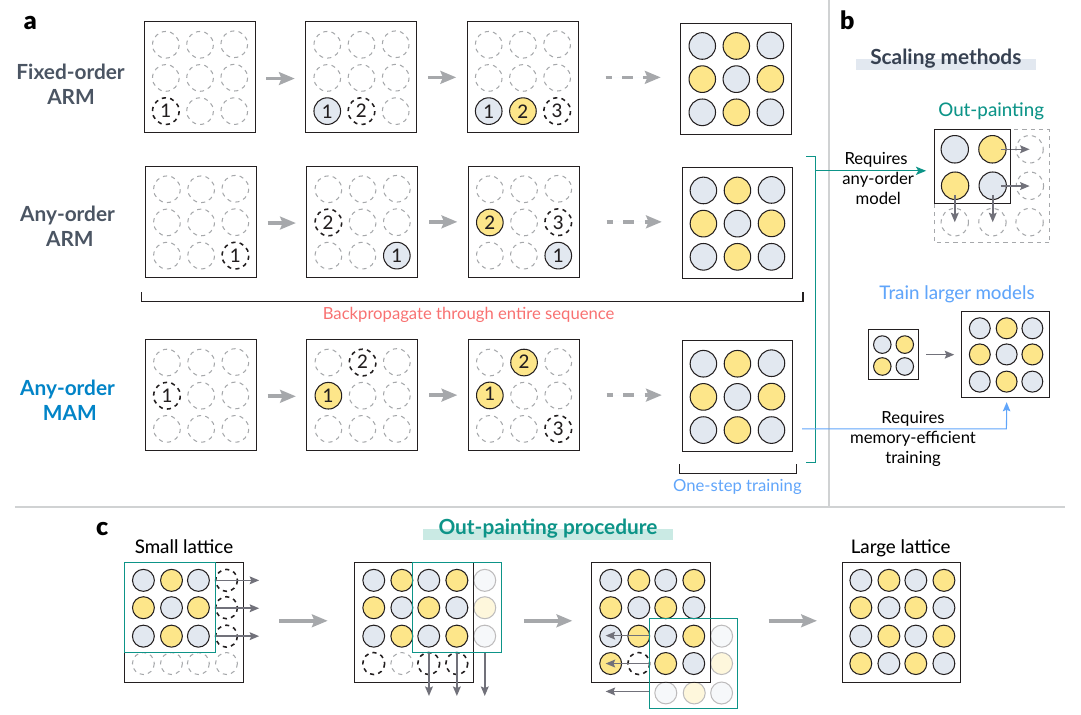}
    \caption{Overview of autoregressive (ARM) and marginalization-based (MAM) frameworks for scaling generative models of lattice thermodynamics.
    (a) Schematic of fixed-order ARM, any-order ARM, and any-order MAM.
    (b) Scaling approaches: out-painting extends smaller core regions; MAMs enable direct training on larger lattices via improved memory efficiency.
    (c) Schematic illustration of the out-painting procedure.}
    \label{fig:method_overview}
\end{figure*}

\subsection{Method development}
\label{sec:methods_conceptual}
Fig.~\ref{fig:method_overview} illustrates our modeling framework combining any-order autoregressive models (AO-ARMs) and MAMs for lattice thermodynamics. This framework enables two complementary scaling strategies (Fig.~\ref{fig:method_overview}b): out-painting from smaller trained models to generate larger lattices, and memory-efficient direct training of MAMs on larger systems.

\subsubsection{Any-order autoregressive models for lattice thermodynamics}
\label{subsec:fo_arm}
ARMs approximate the Boltzmann distribution $\pi(\bx)$ over lattice configurations $\bx = (x_1,\dots,x_L)$ by modeling the generation process as a sequence, where each lattice site is sampled conditioned on all previously generated sites. This sequential factorization is expressed as $\pp(\bx) = \prod_{\ell=1}^{L} \pp(x_\ell \mid \bx_{<\ell})$, where $\pp$ is parameterized by a neural network. Prior work on lattice thermodynamics~\cite{wu_solving_2019, nicoli_asymptotically_2020, damewood_sampling_2022} has used FO-ARMs with predetermined orderings, trained by minimizing the Kullback--Leibler (KL) divergence to $\pi(\bx)$ via the REINFORCE algorithm~\cite{williams_simple_1992} (see Methods: Autoregressive models for details). However, FO-ARMs (Fig.~\ref{fig:method_overview}a, top) face critical scaling limitations: sequential sampling requires backpropagation through the entire sequence, with \emph{O}($L$) cost per lattice site, resulting in \emph{O}($L^2$) computational cost per training step for all $L$ sites. Additionally, the rigid ordering prevents arbitrary conditional generation needed for out-painting to larger lattices.

We introduced AO-ARMs to lattice thermodynamics (Fig.~\ref{fig:method_overview}a, middle). The key idea is to train the model across all possible site orderings simultaneously, so that it learns to predict any site given any subset of already-determined sites. Formally, AO-ARMs represent conditionals $\pp\big(x_{\sigma(\ell)} \,\big|\, \bx_{\sigma(<\ell)}\big)$ along arbitrary permutations $\sigma$ of lattice sites: during training, we sampled random permutations $\sigma$ and trained the model to predict $x_{\sigma(\ell)}$ conditioned on $\bx_{\sigma(<\ell)}$, encouraging consistent conditional distributions regardless of ordering. This enables arbitrary masking patterns and out-painting procedures (Fig.~\ref{fig:method_overview}b--c) where we iteratively filled boundary regions around a core lattice while maintaining approximate equilibrium with the Boltzmann distribution.

\subsubsection{Marginalization models for lattice scaling}
\label{subsec:mam}
While AO-ARMs enable flexible conditioning, they still require sequential evaluation of all $L$ sites per sample, limiting training scalability. To address this bottleneck, we introduced MAMs to lattice thermodynamics (Fig.~\ref{fig:method_overview}a, bottom). MAMs approximate marginals $p(\bx_\mathcal{S}) = \sum_{\bx_{\mathcal{S}'}} p(\bx_\mathcal{S}, \bx_{\mathcal{S}'})$ over arbitrary subsets $\mathcal{S}$ of sites and $\mathcal{S}'$ of the complement sites. We parameterized $\pt(\bx_\mathcal{S})$ with a neural network that outputs log-probabilities for partially or fully specified configurations in a single forward pass, as opposed to the sequential evaluation over each site required for ARMs.

Adapting the approach of~\citet{liu_generative_2024} to lattice thermodynamics, we jointly trained $\pt$ and $\pp$ by combining an energy-based KL divergence objective with a consistency loss that enforces alignment between the predicted conditional and marginal likelihoods, $p_\theta(\bx_{\sigma(<\ell)}) p_\phi(x_{\sigma(\ell)} \mid \bx_{\sigma(<\ell)}) = p_\theta(\bx_{\sigma(\le \ell)})$, for random permutations $\sigma$ (see ``Methods: Machine learning models'' for detailed equations). Crucially, MAMs enable one-step training (Fig.~\ref{fig:method_overview}a) where $\pt$ serves as a surrogate for $\pp$ in energy-based objectives, allowing partial masking instead of full-sequence sampling. This \emph{O}($L$) reduction in training compute and memory costs compared to ARMs enables direct training on larger lattices (Fig.~\ref{fig:method_overview}b, bottom) and the use of more expressive architectures like graph-neural networks (GNNs) and Transformers.

\subsection{Ising model results}
\label{subsec:ising_results}

\begin{figure*}
    \centering
    \includegraphics[width=\textwidth]{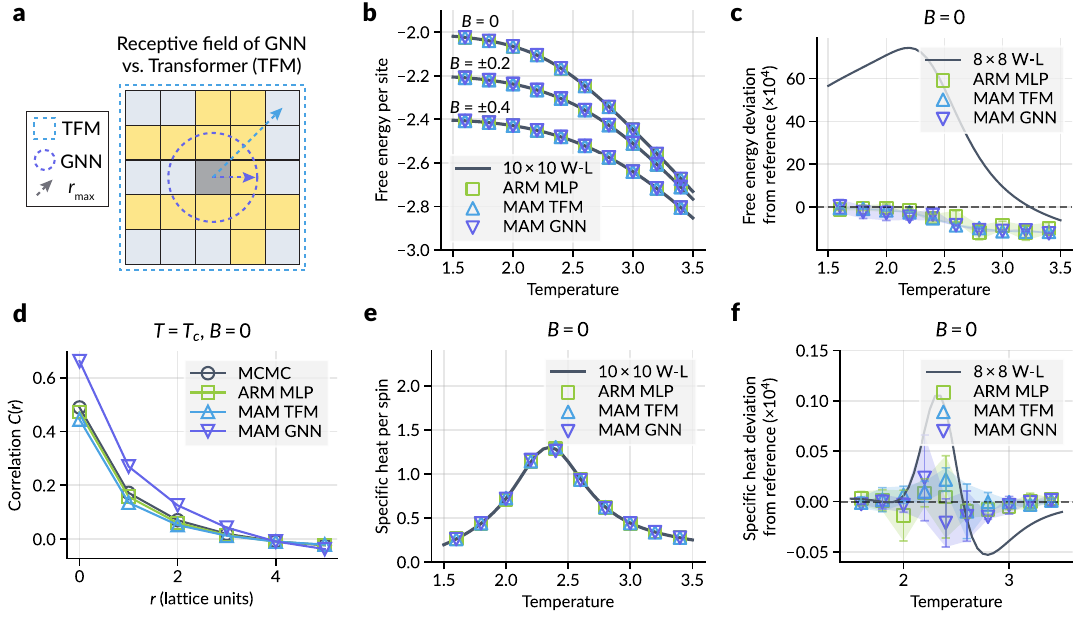}
    \caption{Architectural comparison on the $10 \times 10$ Ising model.
    (a) Schematic comparison of GNN and Transformer (TFM) architectures applied to lattice sites.
    (b--c) Free energy per site and deviation from $10\times10$ Wang--Landau reference at $B=0$. Deviations are multiplied by $10^4$ for visual clarity.
    (d) Spin--spin correlation functions at $T = T_c$ and $B=0$.
    (e--f) Specific heat capacity per site and deviation from $10\times10$ Wang--Landau reference also at $B=0$.}
    \label{fig:ising_architectures}
\end{figure*}

We first benchmark our framework on the two-dimensional Ising model, which provides a controlled setting with known critical behavior and reference free energies. We focus on square lattices of sizes $10\times10$, $15\times15$, and $20\times20$ with periodic boundary conditions, and study the dependence of model performance on architecture, lattice size, and on out-painting vs. training. We implement ARMs and MAMs using three neural architectures—MLPs, GNNs, and Transformers—with Transformers offering global self-attention to capture the long-range correlations near critical points (Fig.~\ref{fig:ising_architectures}a; Methods: Neural network architectures).

\subsubsection{Architectural comparison on $10\times10$ Ising}
\label{subsec:ising_architectures}

To establish baseline performance, we trained AO-ARM and MAM models on the $10 \times 10$ Ising system across a grid of temperatures and external fields. We compared MLP-based AO-ARMs with GNN and Transformer-based MAMs at similar training times (Fig.~\ref{fig:ising_architectures}a), referencing Wang--Landau sampling results as appropriate. GNNs use local message-passing layers that aggregate information from neighboring sites, offering explicit spatial locality but with limited receptive field depth. In contrast, our implementation of Transformer attention layers enables global coupling of all lattice sites, with periodic positional encodings (Methods: Periodic positional embeddings for Transformers) designed to respect lattice boundary conditions, essential for capturing long-range correlations near critical temperatures. MLPs treat all sites equally without explicit spatial structure encoding. For each model, we evaluated the free energy, heat capacity, and spin--spin correlation functions (representative samples are shown in Fig.~S1).

Fig.~\ref{fig:ising_architectures}b shows that all architectures achieve comparable free energies at zero field ($B=0$), closely matching the Wang--Landau reference across the entire temperature range. Fig.~\ref{fig:ising_architectures}c further shows that deviations of the learned free energies from the $10 \times 10$ Wang--Landau reference are mostly smaller than the deviation between the $8 \times 8$ and $10 \times 10$ Wang--Landau results themselves, highlighting the accuracy of the learned models relative to finite-size reference effects. However, differences emerge when examining spin--spin correlation functions. The full correlation analysis across temperatures and fields (Fig.~S2) reveals that at and below the critical temperature at $B=0$ (Fig.~\ref{fig:ising_architectures}d), Transformer MAMs correctly reproduce the expected long-range decay behavior, while GNN MAMs exhibit prematurely flattened correlation profiles due to their limited receptive field depth. ARM MLPs also perform well in capturing correlations despite lacking explicit spatial structure encoding. 

For specific heat capacity (Fig.~\ref{fig:ising_architectures}e--f), all models reproduce the characteristic peak near $T_c$, though MAM GNN exhibits the largest variance and deviation from the reference, while ARM MLP also shows relatively high variance. Examining the interaction energy and up-spin concentration distributions (Figs.~S3--S4) reveals a critical failure mode: GNN MAMs suffer from mode collapse at low temperatures and zero field ($B=0$), failing to properly explore the configuration space. This behavior is consistent with the limited receptive field preventing the GNN model from learning long-range ordered states.

The effective sample size (ESS) analysis (Fig.~S10a--c) further quantifies sampling efficiency across the temperature--field grid. MAM Transformers achieve the highest ESS across most conditions, indicating robust sampling. MAM GNNs perform reasonably well except at and below the critical temperature at $B=0$, where the mode collapse significantly degrades sampling efficiency. ARM MLPs show reduced ESS at higher temperatures, reinforcing the superior performance of Transformer-based architectures for lattice thermodynamics.

While ARM MLPs offer faster inference for small systems and GNN MAMs provide explicit spatial locality, Transformer-based MAMs excel in capturing long-range correlations and achieving better sampling efficiency, which are critical advantages for scaling to larger lattices.

\begin{figure*}
    \centering
    \includegraphics[width=\textwidth]{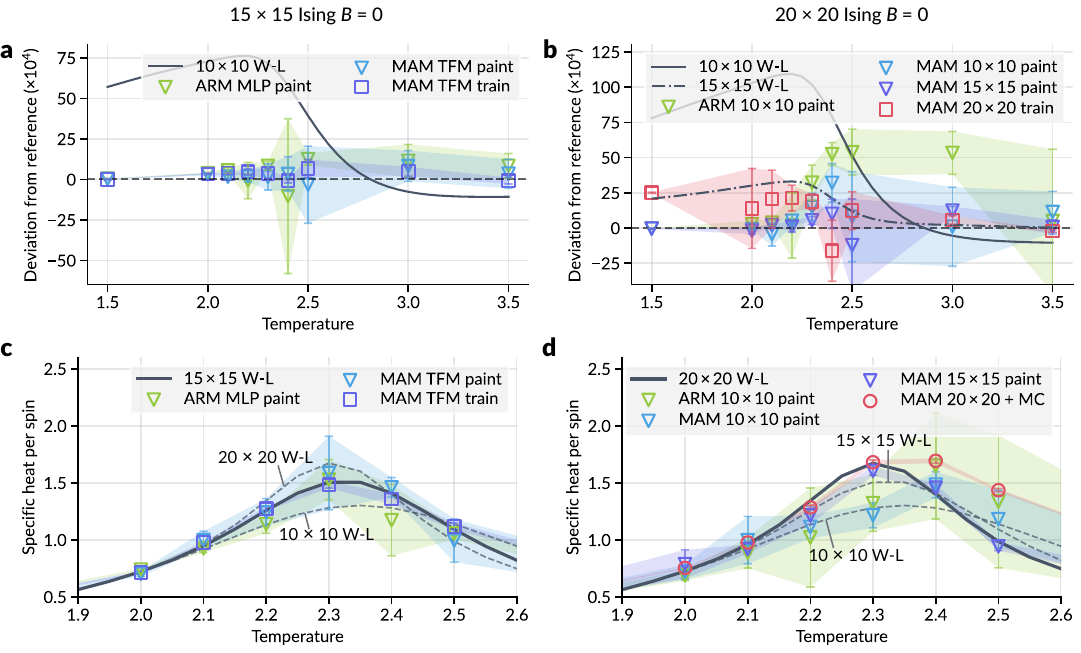}
    \caption{Out-painting and directly-trained results using various ARM MLP and MAM Transformer models on $15\times 15$ and $20\times 20$ Ising lattices at $B = 0$ and across temperatures. 
    (a-b) Free-energy per site deviations from same-sized Wang--Landau reference (with cross-lattice-size comparisons). Values are multiplied by $10^4$ for visual clarity.
    (c-d) Specific heat capacity per site referenced to same-sized Wang--Landau energies.}
    \label{fig:ising_scaling}
\end{figure*}

\subsubsection{Scaling to $15\times15$ and $20\times20$ Ising with MAM Transformers}
\label{subsec:ising_scaling}

Having established the performance of Transformer-based MAMs on $10\times10$ lattices, we examined two approaches for scaling to larger systems: out-painting from smaller trained models to generate $15\times15$ and $20\times20$ configurations (Methods: Out-painting details), and direct training of MAMs on the larger lattices, which is not practical for ARMs due to memory constraints. 

For the $15\times15$ system (Fig.~\ref{fig:ising_scaling}a), both out-painted MAM Transformers from $10\times10$ and directly-trained MAM Transformers on $15\times15$ show small deviations from the Wang--Landau reference free energy at $B=0$. Out-painted ARM MLPs exhibit larger deviations and higher variance, particularly above the critical temperature. Nevertheless, the absolute free energy values (Fig.~S5a) show close agreement between different training and out-painting strategies. The specific heat capacity curves (Fig.~\ref{fig:ising_scaling}c) show that all models capture the characteristic peak near $T_c$, with directly-trained MAM Transformers achieving the best agreement and lowest variance at higher temperatures, followed by out-painted MAM Transformers (Fig.~S5c). The directly-trained $15\times15$ model also exhibits the highest ESS across the temperature--field grid compared to out-painted models (Fig.~S10d--f), indicating more efficient sampling.

For the $20\times20$ system (Fig.~\ref{fig:ising_scaling}b,d), the performance gap between approaches becomes more pronounced. Out-painted models from $10\times10$ show larger deviations than those from $15\times15$ (Fig.~\ref{fig:ising_scaling}b, Fig.~S5d), consistent with cumulative approximation errors from smaller training systems. The directly-trained MAM Transformer on $20\times20$ achieves the closest agreement with reference free energies at temperatures above $T_c$ (Fig.~\ref{fig:ising_scaling}b), but exhibits larger deviations near and below the critical temperature, where precisely learning both up and down spin modes becomes challenging (Fig.~S9). Unlike GNN MAMs at smaller system sizes, which suffer from complete mode collapse, Transformer-based models recognize that there are roughly two modes; however, the interaction energy and up-spin concentration distributions (Figs.~S6--S9) reveal that they do not successfully mirror the target statistics at these challenging conditions. 

ESS examination reveals important limitations of direct training alone (Fig.~S10g--j). While the directly-trained $20\times20$ model shows higher ESS than out-painted models from $10\times10$ at higher temperatures, it exhibits limited ESS at low temperatures and extreme field values, with normalized effective sample size (NESS) values often below 0.5 in these regions. In contrast, out-painting from $15\times15$ models achieves consistently higher ESS across a broader range of thermodynamic conditions, particularly at low to moderate temperatures, making it the most robust approach for the $20\times20$ system. When the directly-trained $20\times20$ samples are used as starting points for MCMC refinement (short MCMC chains initialized from model-generated configurations to correct residual errors in the learned distribution), ESS improves dramatically across all conditions (Fig.~S10k), and the heat capacity near and below $T_c$ improves significantly (Fig.~\ref{fig:ising_scaling}d, Fig.~S5d). However, despite improved sample quality, free-energy accuracy lags behind that of out-painted samples from $15\times15$ models in the $2.4 < k_\text{B}T < 3.0$ range.

The choice between direct training and out-painting depends on the specific application requirements. Currently, out-painting from $15\times15$ models provides the best balance of accuracy, sampling efficiency, and computational cost for $20\times20$ systems, requiring no additional training and achieving robust performance across the temperature--field grid. Direct training of MAM Transformers on $20\times20$ lattices remains computationally feasible where fixed-order ARMs become memory-prohibitive, but the limited ESS at low temperatures and near $T_c$ limits its effectiveness without MCMC refinement. For systems beyond $20\times20$, out-painting becomes the primary viable approach, enabling our framework to scale to larger lattices where traditional ARM MLPs are impractical.

\subsection{CuAu alloy results}
\label{sec:cuau_results}

\begin{figure*}
    \centering
    \includegraphics[width=\textwidth]{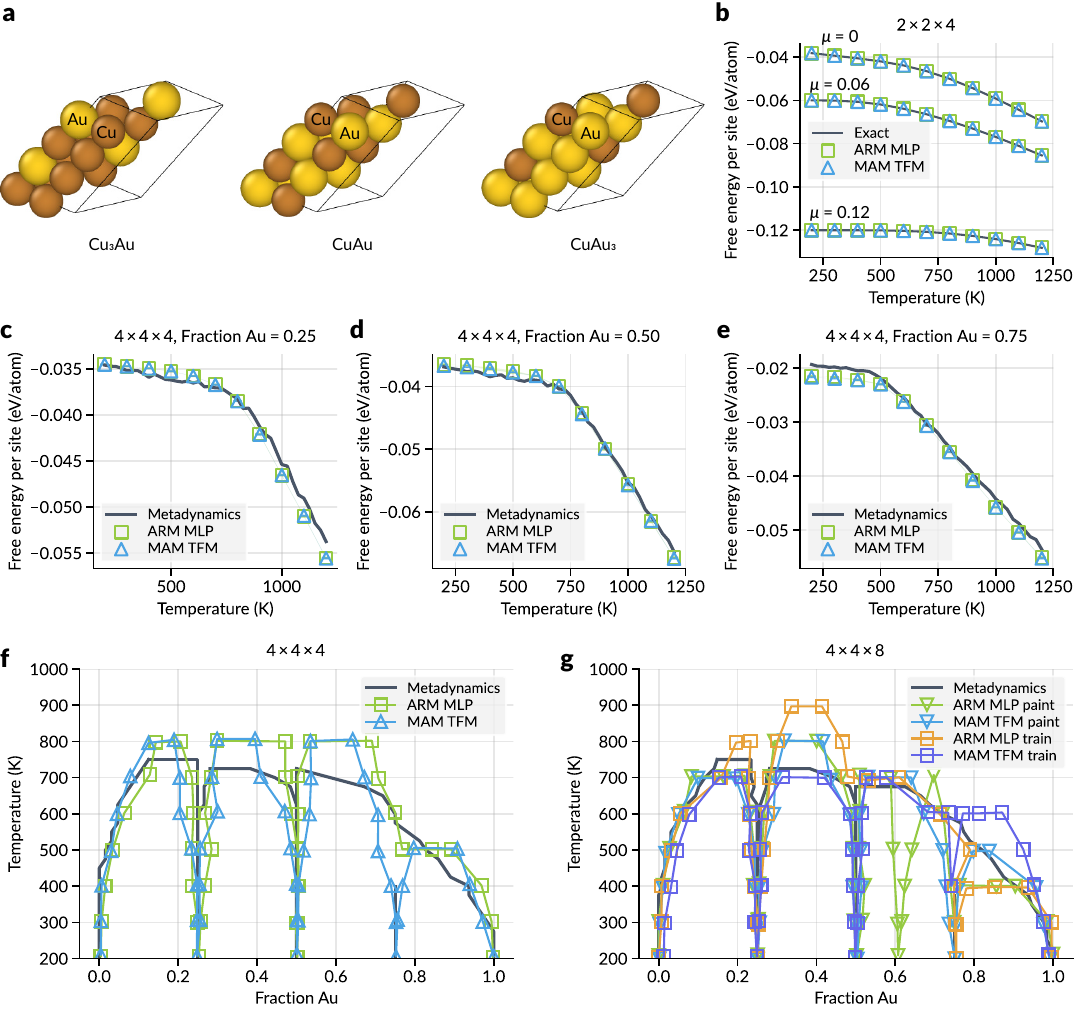}
    \caption{CuAu alloy results. 
    (a) Cu-Au ordered intermetallic phases observed. From left to right: (1) $\ce{Cu3Au}$, (2) CuAu, and (3) $\ce{CuAu3}$.
    (b) Free-energy per site comparison between models trained on the $2\times2\times4$ supercell and exact results from enumeration.
    (c-e) Predicted free-energy per site for the $4\times4\times4$ supercell at specific compositions compared with metadynamics reference.
    (f-g) Predicted temperature-composition phase diagram for $4\times4\times4$ and $4\times4\times8$ supercells compared with metadynamics reference.}
    \label{fig:cuau_phase}
\end{figure*}

We next applied our framework to a more realistic materials system: the CuAu alloy described by cluster expansion. Unlike the Ising model, which has only two ordered phases, the CuAu system exhibits multiple ordered alloy phases at low temperatures (e.g., $\ce{Cu3Au}$, $\ce{CuAu}$, and $\ce{CuAu3}$, see Fig.~\ref{fig:cuau_phase}a) and a disordered solid-solution phase at high temperatures, providing a rigorous test of our models' ability to capture complex phase behavior.

\subsubsection{$2\times2\times4$ CuAu baseline}
\label{subsec:cuau_baseline}
We first trained AO-ARM and MAM models on a $2\times2\times4$ CuAu supercell, following the cluster-expansion setup of Ref.~\citenum{damewood_sampling_2022} (see Methods). The thermodynamics were explored over a grid of temperatures and Au chemical potentials.

At this small supercell size, exact free energies are computed by enumerating all $2^{16} = 65,536$ possible configurations (Methods: CuAu alloy exact free energy calculations), providing a precise benchmark. Both ARM MLPs and MAM Transformers achieve nearly perfect agreement with exact free energies across all temperatures and chemical potentials tested (Fig.~\ref{fig:cuau_phase}b), with curves that are visually indistinguishable from the exact reference. Quantitatively, MAM Transformers show slightly lower average deviations from exact values (Fig.~S11a). The interaction energy distributions and Au concentration distributions (Figs.~S12--S13) confirm that both models successfully reproduce the target statistics across all thermodynamic conditions, with all ordered intermetallic phases ($\ce{Cu3Au}$, $\ce{CuAu}$, and $\ce{CuAu3}$) well-represented with respect to the MCMC reference. Both models also exhibit high ESS across the temperature-chemical potential grid (Fig.~S19a--b), indicating efficient sampling.

\subsubsection{Scaling to $4\times4\times4$ CuAu via training}
\label{subsec:cuau_scaling}
To explore larger supercells, we trained both ARM MLPs and MAM Transformers on the $4\times4\times4$ CuAu system. Free energies at specific compositions were evaluated by binning samples and computing weighted averages (Methods: Observables calculations). At Au fraction $x_\text{Au} \in \{0.25, 0.50, 0.75\}$, both models show good agreement with the metadynamics reference across the temperature range (Fig.~\ref{fig:cuau_phase}c--e, Fig.~S11b--c). Deviations are minimal at low to moderate temperatures and increase to at most 10 meV/atom at the highest temperatures ($T > 1000$ K), where the models slightly overestimate the free energy relative to metadynamics. MAM Transformers exhibit higher ESS across the $T$--$\mu$ grid (Fig.~S19c--d), and their sampled interaction energies are more closely aligned with the MCMC reference (Fig.~S14), indicating more efficient sampling.

We also constructed temperature--composition phase diagrams from model samples and compared the predicted two-phase coexistence boundaries with those obtained from metadynamics (Fig.~\ref{fig:cuau_phase}f, Fig.~S18a--b, details in Methods: Observables calculations). Both models successfully capture the overall shape of the ordered phase region, which forms a dome-like structure with maximum transition temperatures around 700--750 K for Au fractions between 0.2 and 0.8. However, important qualitative differences emerge: the ARM MLP model fails to fully capture the $\ce{CuAu3}$ phase at low temperatures, missing a significant portion of the phase diagram. In contrast, the MAM Transformer model reproduces all three ordered phases ($\ce{Cu3Au}$, $\ce{CuAu}$, and $\ce{CuAu3}$), though with slightly broader coexistence lines that indicate more variance in the predicted phase boundaries. 

\subsubsection{Scaling to $4\times4\times8$ CuAu via out-painting and direct training}
\label{subsec:cuau_scaling_outpainting}
For the $4\times4\times8$ system, we compared two scaling strategies: out-painting from $4\times4\times4$ models and direct training on the $4\times4\times8$ supercell (Methods: Out-painting details). This comparison more clearly demonstrates the advantages of MAM Transformers and out-painting than in the Ising case, where the trade-offs were more nuanced.

Free energies at specific compositions (Fig.~S11d--h) show that the directly-trained MAM Transformer samples and the out-painted ARM and MAM samples achieve strong agreement with the metadynamics reference across most compositions and temperatures, with deviations typically below 5 meV/atom. An exception is observed for the directly-trained MAM Transformer at $x_\text{Au} = 0.75$ and $T = 200 \text{K}$ (Fig.~S11g), where deviations are larger. In contrast, the directly-trained ARM MLP shows significant deviations at low temperatures for the $\ce{CuAu}$ composition ($x_\text{Au} = 0.50$), with errors approaching 40 meV/atom. Effective sample sizes are comparable across all models (Fig.~S19e--h), indicating that sampling efficiency is not the primary differentiator; rather, the accuracy of the learned distributions distinguishes the models.

The temperature--composition phase diagrams (Fig.~\ref{fig:cuau_phase}g, Fig.~S18c--f) reveal more pronounced differences between approaches. The directly-trained MAM Transformer model most faithfully reproduces the metadynamics phase boundaries, capturing all three ordered intermetallic phases with transition temperatures typically within $\sim 100$ K of the reference. It slightly overpredicts the transition temperature from ordered to disordered phase at high Au fractions ($x_\text{Au} > 0.75$) but otherwise closely follows the metadynamics curve. The out-painted MAM Transformer model performs comparably, with only minor deviations from the directly-trained model, demonstrating that out-painting is highly effective for this system.

In contrast, ARM MLP models exhibit systematic failures. The directly-trained ARM MLP misses the $\ce{CuAu}$ phase entirely at low temperatures (explaining the significant error for free energies at the $x_\text{Au} = 0.50$ composition) and overpredicts the coexistence curves between Cu, $\ce{Cu3Au}$, and $\ce{CuAu}$ phases at high temperatures. The out-painted ARM MLP model captures the $\ce{CuAu}$ phase (unlike the directly-trained version), suggesting that out-painting can sometimes compensate for training limitations, but it also introduces artifacts, including a spurious phase boundary near $x_\text{Au} = 0.6$ containing a mixture of $\ce{CuAu}$ and $\ce{CuAu3}$ phases. The interaction energy and $x_\text{Au}$ distributions (Figs.~S16--S17) confirm that MAM Transformers better represent the full thermodynamic ensemble across all phases.

These results demonstrate that MAM Transformers, particularly when combined with out-painting strategies, provide a robust and accurate approach for exploring CuAu phase diagrams at larger system sizes, with clear advantages over ARM MLPs in capturing the full range of ordered intermetallic phases.

\subsection{Efficiency comparison}
\label{subsec:efficiency}

\begin{figure*}
    \centering
    \includegraphics[width=\textwidth]{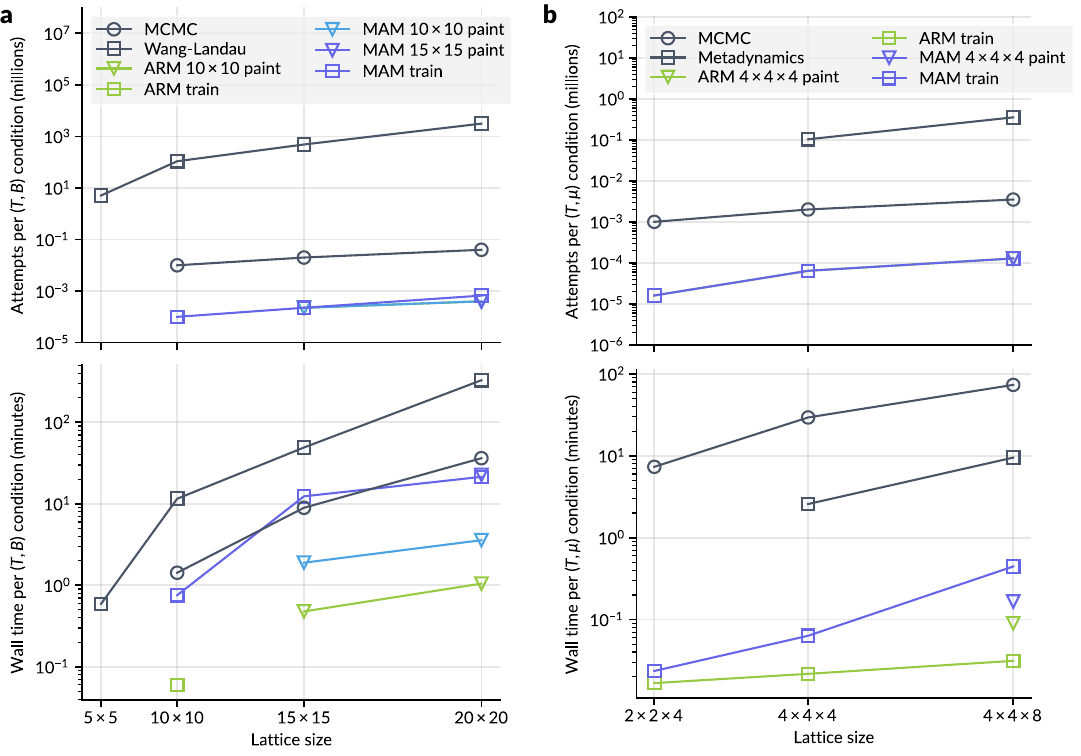}
    \caption{Speed comparison between AO-ARM MLPs and AO-MAM Transformers.
    (a) Number of attempts and wall-clock time per $(T, B)$ condition for various Ising model lattice sizes comparing trained and out-painted samples with MCMC and Wang--Landau reference sampling. Note for Wang--Landau, the wall-clock time is per $B$ condition.
    (b) Number of attempts and wall-clock time per $(T, \mu)$ condition for various CuAu model lattice sizes comparing trained and out-painted samples with MCMC and metadynamics reference sampling. Note for metadynamics, the wall-clock time is per $T$ condition.}
    \label{fig:efficiency}
\end{figure*}

Finally, we performed a systematic comparison of computational costs between traditional sampling methods (MCMC, Wang--Landau, metadynamics) and our learned generative models (AO-ARM MLPs and AO-MAM Transformers), considering both training and sampling phases (Fig.~\ref{fig:efficiency}). The learned models require an upfront training investment: for example, $20\times20$ Ising MAM Transformers require 30 hours on four GPUs, while $4\times4\times8$ CuAu MAM Transformers require 60 hours on a single GPU (Methods: Training details). In terms of wall-clock sampling time, for $20\times20$ Ising systems, MAM Transformers ($\sim 30$ minutes per $(T, B)$ condition) are broadly comparable to MCMC ($\sim 40$ minutes per $(T, B)$ condition), while Wang--Landau requires $\sim 300$ minutes per $B$ condition. For $4\times4\times8$ CuAu systems, the wall-clock advantage is more pronounced: MAM Transformers require $\sim 0.5$ minutes per $(T, \mu)$ condition compared to $\sim 80$ minutes for MCMC and $\sim 10$ minutes for metadynamics (the latter per $T$ condition).

The efficiency gains are also reflected in the number of sampling attempts required per condition (Fig.~\ref{fig:efficiency}). For $20\times20$ Ising systems, MAM Transformers require $400$ attempts (one attempt per site) compared to $\sim 10^9$ for Wang--Landau and $\sim 5 \times 10^4$ for MCMC. For $4\times4\times8$ CuAu systems, MAM Transformers require $128$ attempts (also one attempt per site) compared to $\sim 5 \times 10^5$ for metadynamics and $\sim 5 \times 10^3$ for MCMC. Once trained, our models generate samples without requiring additional energy evaluations. While the relationship between the number of attempts and wall-clock time depends on the computational cost per attempt (which varies with hardware and implementation), the combination of orders-of-magnitude reduction in attempts and the elimination of per-sample energy evaluations provides a significant advantage for large-scale simulations. This suggests that further improvements in hardware or sampling algorithms could yield even greater speedups for the learned models.

A key advantage of the learned models is that a single training run produces a model that generalizes across the entire thermodynamic grid: at inference, each $(T, B)$ or $(T, \mu)$ condition requires only a fast forward pass with no retraining or additional energy evaluations, in contrast to MCMC and metadynamics, which must be rerun from scratch for every new condition. The one-time training cost is therefore quickly amortized when many conditions must be sampled, as is typical in phase diagram mapping. Out-painting strategies further improve efficiency by enabling exploration of larger lattices without additional training: for instance, out-painting from $15\times15$ to $20\times20$ Ising or from $4\times4\times4$ to $4\times4\times8$ CuAu requires no retraining and achieves sampling times similar to or even lower than directly-trained models. ARM MLPs offer faster training and sampling than MAM Transformers (e.g., $4\times4\times8$ CuAu: 12 hours training and $\sim 0.03$ minutes sampling per condition for ARM vs. 60 hours training and $\sim 0.5$ minutes sampling per condition for MAM), but MAM Transformers provide superior accuracy and better capture of complex phase behavior, making them preferable when accuracy is prioritized over raw speed.

\section{Discussion}
\label{sec:discussion}
Our results demonstrate three key advances for scalable lattice thermodynamics. First, Transformer-based MAMs outperform MLP and GNN variants on the same lattice size by capturing long-range correlations near critical conditions and phase transitions, with GNNs limited by finite receptive fields. Second, MAMs enable scaling beyond fixed-order ARMs through memory-efficient training that facilitates deeper architectures on larger lattices. We successfully scaled AO-MAM Transformers to $15\times15$ and $20\times20$ Ising lattices under realistic hardware constraints where fixed-order ARMs become impractical. For CuAu systems, MAM Transformers capture all three ordered intermetallic phases ($\ce{Cu3Au}$, $\ce{CuAu}$, and $\ce{CuAu3}$) on the $4\times4\times4$ supercell, where ARM MLPs miss the $\ce{CuAu3}$ phase, and achieve free energy deviations below 5 meV/atom on the $4\times4\times8$ supercell with phase boundaries within $\sim 100$ K of the metadynamics reference. Third, any-order training enables out-painting that expands smaller models to larger lattices with thermodynamic accuracy comparable to direct training, while requiring no additional training. Notably, out-painting from $15\times15$ to $20\times20$ Ising using MAM Transformers achieves consistently higher ESS across broader thermodynamic ranges than direct training alone (Fig.~S10i--j), demonstrating that out-painting can outperform direct training when source models are sufficiently large and well-trained.

Several limitations and opportunities for improvement remain. Out-painted samples can exhibit thermodynamic distributions that differ from those of directly-trained models on the target lattice size (Figs.~S6--S7, S16--S17). These effects can be mitigated through importance sampling, though at the cost of additional samples and energy evaluations. For directly-trained larger models (e.g., $20\times20$ Ising), reduced ESS at low temperatures and near criticality could be addressed through alternative training objectives that provide more gradient information per sample~\cite{zhu_mdns_2025}, training schedules that reduce distribution shift over consecutive epochs~\cite{guo_proximal_2025}, or hybrid approaches that combine learning with targeted MCMC refinement in challenging regions of the phase space.

Looking forward, these capabilities open several avenues for applications. For surface and interface systems, AO-ARM and MAM models could generate reconstructions and adsorbate configurations conditioned on experimental observations or partial structural information. For complex alloys and disordered oxides, they may help explore phase diagrams and disorder effects in supercells that are challenging to access via direct MCMC or first-principles calculations. A particularly promising direction is the extension to off-lattice systems, where discrete elemental identities are combined with continuous atomic positions. Coupling our discrete lattice framework with continuous coordinate models could enable efficient exploration of realistic material systems where both discrete ordering and continuous relaxation contribute to thermodynamic behavior~\cite{wexler_automatic_2019, xu_atomistic_2022,du_machine-learning-accelerated_2023, du_accelerating_2025}. Finally, integrating our framework with modern MLFFs, either for energy-based sampling or fine-grained evaluation, could ultimately enable autonomous, thermodynamics-aware generative pipelines for materials design.

\section{Methods}

\subsection{Overview of lattice thermodynamics and semigrand ensembles}
We consider lattice systems described by discrete site variables 
$\mathbf{x} = (x_1, \ldots, x_L)$, where each site $x_\ell$ takes values in a 
finite set representing spin or atomic species. 
The equilibrium distribution is given by the semigrand canonical ensemble:
\begin{align}
    p(\mathbf{x}) = \frac{1}{Z} 
    \exp\!\left[
        -\frac{E(\mathbf{x}) - \sum_i \mu_i N_i(\mathbf{x})}{k_{\mathrm{B}}T}
    \right],
\end{align}
where $Z$ is the partition function, $E(\mathbf{x})$ is the lattice energy, $\mu_i$ is the chemical potential for species~$i$, $N_i(\mathbf{x})$ counts the number 
of species~$i$, $k_\text{B}$ is the Boltzmann constant, and $T$ is the temperature.

\subsection{Machine learning models}
\subsubsection{Autoregressive models}\label{subsec:methods_arm}
ARMs factorize the target Boltzmann distribution $\pi(\mathbf{x})$ according to an ordering of lattice sites. Let $\sigma$ denote a permutation of $\{1,\dots,L\}$ that specifies the generation order. The joint distribution is factorized as~\cite{wu_solving_2019, nicoli_asymptotically_2020, damewood_sampling_2022}:
\begin{align}
    p_\phi(\mathbf{x}) 
    = \prod_{\ell=1}^L p_\phi\!\left(x_{\sigma(\ell)} \mid \mathbf{x}_{\sigma(<\ell)}\right),
\end{align}
where $p_\phi$ is parameterized by a neural network with parameters $\phi$, and $\mathbf{x}_{\sigma(<\ell)}$ denotes the sites generated before step $\ell$ according to ordering $\sigma$.

We distinguish between two training strategies. \textbf{FO-ARMs} use a single fixed permutation $\sigma$ (typically the identity), leading to a deterministic generation order but potential ordering bias. \textbf{AO-ARMs} train by sampling a random permutation~$\sigma$ at each training step, learning $p_\phi(x_{\sigma(\ell)} \mid \mathbf{x}_{\sigma(<\ell)})$ for all possible orderings.

We learn $p_\phi$ by minimizing its KL divergence to the target Boltzmann distribution:
\begin{widetext}
\begin{align}
    \min_\phi D_{\mathrm{KL}}\!\left(p_\phi(\mathbf{x}) \,\Vert\, \pi(\mathbf{x})\right)
    = \min_\phi \mathbb{E}_{\mathbf{x} \sim p_\phi} 
    \left[ \beta\left(E(\mathbf{x}) - \sum_i \mu_i N_i(\mathbf{x})\right) + \log p_\phi(\mathbf{x}) + \log Z\right],
\end{align}
\end{widetext}
where $\beta = 1/(k_{\mathrm{B}}T)$. This is equivalent to minimizing a variational free energy $F_q = \mathbb{E}_{\mathbf{x} \sim p_\phi} \left[ E(\mathbf{x}) - \sum_i \mu_i N_i(\mathbf{x}) + k_\text{B}T \log p_\phi(\mathbf{x}) \right]$. See SI: Variational free energy for more details.

\subsubsection{Marginalization models}
\label{subsec:methods_mam}
MAMs approximate marginals over arbitrary subsets $\mathcal{S} \subseteq \{1,\dots,L\}$ of lattice sites. For a subset $\mathcal{S}$ and its complement $\mathcal{S}'$, the target marginal is:
\begin{align}
    p(\mathbf{x}_\mathcal{S}) = \sum_{\mathbf{x}_{\mathcal{S}'}} p(\mathbf{x}_\mathcal{S}, \mathbf{x}_{\mathcal{S}'})
\end{align}

We parameterize a marginal model $p_\theta(\mathbf{x}_\mathcal{S})$ with a neural network (parameters $\theta$) that takes a configuration that could be partially specified or fully specified (e.g., $\mathbf{x}_\mathcal{S}$ or $\mathbf{x}$) and outputs its log-probability or log-marginal. Following Ref.~\citenum{liu_generative_2024}, we jointly train: (i) a marginal model $p_\theta$ approximating $\pi(\mathbf{x})$, and (ii) an AO autoregressive model $p_\phi$. Their consistency is enforced via a marginalization loss. For a permutation $\sigma$ and index $\ell$:
\begin{widetext}
    \begin{align}
        \mathcal{L}_{\mathrm{Mar}}(\mathbf{x}, \sigma, \ell)
        &= 
        \left[
        \log p_\theta\!\left(\mathbf{x}_{\sigma(\le \ell)}\right)
        -
        \log p_\theta\!\left(\mathbf{x}_{\sigma(< \ell)}\right)
        -
        \log p_\phi\!\left(x_{\sigma(\ell)} \mid \mathbf{x}_{\sigma(<\ell)}\right)
        \right]^2,
    \end{align}
\end{widetext}

which penalizes deviations from the marginalization identity $p(\mathbf{x}_{\sigma(<\ell)}) p(x_{\sigma(\ell)} \mid \mathbf{x}_{\sigma(<\ell)}) = p(\mathbf{x}_{\sigma(\le \ell)})$.

The joint training objective combines the KL divergence between $p_\theta$ and the target Boltzmann distribution with the consistency term:
\begin{align}
    \min_{\theta,\phi}
    \ 
    D_{\mathrm{KL}}\!\left(p_\theta(\mathbf{x}) \,\Vert\, \pi(\mathbf{x})\right)
    +
    \lambda \, \mathbb{E}_{\mathbf{x},\sigma,\ell}
    \left[ \mathcal{L}_{\mathrm{Mar}}(\mathbf{x},\sigma,\ell) \right],
\end{align}
where $\lambda$ controls the strength of the consistency regularization.

\subsubsection{Training by REINFORCE}
To train the models, we draw a batch of samples, $\bx$, using an autoregressive model (ARM or MAM) and estimate the gradient of the KL divergence with the REINFORCE algorithm~\citep{williams_simple_1992, wu_solving_2019}:

\begin{widetext}
\begin{align*}
    \nabla_\phi D_{\text{KL}}\left( p_\phi(\bx) \left\Vert \frac{f(\bx)}{Z} \right. \right)
     &= \nabla_\phi \mathbb{E}_{\bx \sim \pp(\bx)} [ \log \pp(\bx) - \log f(\bx)] \qquad \text{(from SI: Variational free energy)} \\
    &= \mathbb{E}_{\bx \sim p_\phi(\bx)} \left[\nabla_\phi \log \pp (\bx) \left( \log \pp(\bx) - \log f(\bx) \right) \right]  \\
    &\approx \frac{1}{N} \sum_{i=1}^N \nabla_\phi \log \pp (\bx^{(i)}) \left( \log \pp(\bx^{(i)}) - \log f(\bx^{(i)}) \right)
\end{align*}
\end{widetext}

where $N$ is the number of samples in the batch, 
$\bx^{(i)}$ is the $i$-th sample, and $\nabla_\phi$ is the gradient with respect to the parameters of the autoregressive model. Samples are drawn from a range of temperatures and fields (chemical potentials). To reduce variance, the $\log \pp(\bx^{(i)}) - \log f(\bx^{(i)})$ term is standardized within each batch by subtracting the mean and scaling by the standard deviation, computed separately for each $(T, \mu)$ condition.

\subsection{Neural network architectures}
\label{subsec:methods_architectures}

\subsubsection{Multilayer perceptrons}  
MLPs treat lattice sites as flattened inputs and outputs. We use three-layer fully connected networks with width 2048 and residual connections between hidden layers. Temperature and field (chemical potential) are directly concatenated to the one-hot encoded lattice configuration. The number of parameters scales as $\mathcal{O}(L \cdot K)$ for $K$ site types (e.g., $K=2$ for Ising and CuAu), limiting their practicality for large supercells.

\subsubsection{Graph neural networks}
We adopted the equivariant PaiNN model~\citep{schutt_equivariant_2021} as the GNN architecture. PaiNN incorporates scalar and vector node features, with the structure encoded using a radial basis function expansion of atomic distances and unit vector directions along edges. We embed temperature and chemical potential using sinusoidal Fourier features~\citep{vaswani_attention_2023} and add them to the atom embeddings. Our implementation uses 3 message-passing layers with 64 feature dimensions and 20 radial basis functions. Edges are constructed within a cutoff radius of 2.5~\AA{} with an offset of 0.5~\AA{}. The offset was adopted from~\citet{nam_flow_2025}.

\subsubsection{Transformer-based models}  
Transformers use self-attention mechanisms to couple all sites within a layer, providing a large effective receptive field. We introduce a lattice transformer architecture with: (i) periodic sinusoidal positional embeddings with frequencies quantized to respect lattice periodicity, applied independently to each lattice dimension via Rotary Position Embedding (RoPE)~\cite{su_roformer_2023} (Methods: Periodic positional embeddings for Transformers), (ii) sinusoidal Fourier feature embeddings~\citep{vaswani_attention_2023} for temperature and chemical potential conditions, and (iii) a three-layer MLP output head with Gaussian Error Linear Unit (GELU) activations. The $10 \times 10$ Ising models use 4 Transformer blocks with 256 embedding and hidden dimensions and 8 attention heads. The $15 \times 15$ and $20 \times 20$ models use 5 Transformer blocks with 512 embedding and hidden dimensions and 16 attention heads. $2 \times 2 \times 4$ CuAu models use 4 Transformer blocks with 256 embedding and hidden dimensions and 8 attention heads while $4 \times 4 \times 4$ use 5 Transformer blocks with 256 embedding and hidden dimensions and 16 attention heads. $4 \times 4 \times 8$ CuAu models use 5 Transformer blocks with 512 embedding and hidden dimensions and 16 attention heads.

\subsubsection{Periodic positional embeddings for Transformers}
To respect the periodic boundary conditions inherent to lattice systems, we design sinusoidal positional embeddings with frequencies that satisfy periodicity constraints. For a lattice with period $L$ along a given dimension, we require:
\begin{align*}
    \sin(\omega_i x) &= \sin(\omega_i (x+L))\\
    \cos(\omega_i x) &= \cos(\omega_i (x+L))
\end{align*}
for all positions $x$. This constraint is satisfied when $L\omega_i = 2 n_i \pi$ for $n_i \in \mathbb{Z}$, yielding quantized frequencies:
\begin{align*}
    \omega_i = \frac{2 n_i \pi}{L}, \quad n_i = 1, 2, 3, \dots
\end{align*}
We apply this construction independently to each lattice dimension (e.g., $x$, $y$, $z$ for 3D systems), ensuring that the positional embeddings respect the full periodic lattice topology. The resulting embeddings are incorporated into the Transformer architecture via RoPE~\cite{su_roformer_2023}, maintaining equivariance under lattice translations.

\subsection{Training details}
\subsubsection{Ising model training details}
ARM MLP models were trained with batch size 1250. All MAM models used consistency loss parameter $\lambda = 100$ to couple the conditional and marginal networks. MAM Transformer models were trained with 10 Gibbs sampling steps per iteration, batch sizes of 500 for $10 \times 10$ systems and 250 for $15 \times 15$ and $20 \times 20$ systems. MAM GNN models used batch size 250.

During training, samples were drawn at each iteration across 5 temperatures and 5 fields, for a total of 25 unique conditions. The conditions were initially uniformly distributed across $\mu \in [-0.4,+0.4]$ and $k_{\text{B}}T \in [1.5, 3.5]$. The change in thermodynamic conditions per iteration was sampled according to $\Delta \mu \sim \mathcal{N}(0, 0.05^2)$ and $\Delta k_{\text{B}}T \sim \mathcal{N}(0, 0.2^2)$. All $10\times10$ models were trained for 15 hours on a single NVIDIA A100 GPU, $15\times15$ MAM Transformer models were trained for 15 hours on four NVIDIA A100 GPUs, and $20\times20$ MAM Transformer models were trained for 30 hours on four NVIDIA A100 GPUs.

\subsubsection{CuAu alloy training details}
ARM MLP models were trained with batch size 2048 for $2 \times 2 \times 4$ CuAu, 900 for $4 \times 4 \times 4$ CuAu, and 1125 for $4 \times 4 \times 8$ CuAu. All MAM models used consistency loss parameter $\lambda = 100$ to couple the conditional and marginal networks. MAM Transformer models were trained with 4 Gibbs sampling steps per iteration, batch sizes of 1024 for $2 \times 2 \times 4$ systems and 450 for both $4 \times 4 \times 4$ and $4 \times 4 \times 8$ systems.

During training, samples were drawn at each iteration across 4 temperatures and 4 fields for $2 \times 2 \times 4$ CuAu, for a total of 16 unique conditions. For $4 \times 4 \times 4$ and $4 \times 4 \times 8$ CuAu, samples were drawn at each iteration across 5 temperatures and 5 fields for a total of 25 unique conditions. The conditions were initially uniformly distributed across $\mu \in$ [$-0.12$ eV, 0.12 eV] and $T \in$ [200 K, 1200 K]. The change in thermodynamic conditions per iteration was sampled according to $\Delta \mu / \text{eV} \sim \mathcal{N}(0, 0.02^2)$ and $\Delta T / \text{K} \sim \mathcal{N}(0, 1^2)$. ARM MLP $2 \times 2 \times 4$ CuAu models were trained for 3 hours on a single NVIDIA A100 GPU while the MAM Transformer model took 12 hours. ARM MLP $4 \times 4 \times 4$ CuAu models were trained for 4 hours using the same hardware while the MAM Transformer model took 20 hours. Finally, ARM MLP $4 \times 4 \times 8$ CuAu models were trained for 12 hours while the MAM Transformer model took 60 hours, also on a single NVIDIA A100 GPU.

\subsection{Out-painting details}
Out-painting extends a model trained on a smaller lattice to generate configurations on a larger lattice by iteratively filling sites using the learned conditional distributions. We first initialize an empty larger lattice (all sites masked). We then iteratively select regions matching the training lattice size (e.g., $10 \times 10$ or $15 \times 15$ for Ising, or extend along one dimension for CuAu) and use the trained model to conditionally sample the masked sites within that region, given the already-filled neighboring sites. Sites are sampled autoregressively within each region according to the learned conditional probabilities. This process continues until all sites are filled, with regions processed in raster scan order to ensure consistent conditioning on previously filled sites.

\subsubsection{$15 \times 15$ Ising model out-painting}
Using $10 \times 10$ trained ARM MLP and MAM Transformer models, we generated $15 \times 15$ lattices via out-painting. A total of 10,000 samples were generated at each condition. Batches of 2500 $15 \times 15$ lattices were initialized as fully masked (all sites undefined) at 5 temperatures and 5 fields uniformly distributed across $\mu \in [-0.4,+0.4]$ and $k_{\text{B}}T \in [1.5, 3.5]$, for a total of 25 unique conditions. $10 \times 10$ sections were positioned with stride 7 in each dimension (yielding overlapping regions), following a raster scan order from top to bottom and left to right while respecting periodic boundary conditions. Within each $10 \times 10$ section, sites were sampled autoregressively using the model's conditional probabilities, with the sampling order randomized within each section to leverage any-order capabilities. When sections overlapped, previously filled sites were used as conditioning information.

\subsubsection{$20 \times 20$ Ising model out-painting}
$20 \times 20$ lattices were generated using a similar approach to the $15 \times 15$ Ising model out-painting. For out-painting from $10 \times 10$ models, we used stride 7 as above. For out-painting from $15 \times 15$ models, we used stride 7 to tile $15 \times 15$ regions across the $20 \times 20$ lattice, with overlapping regions handled as above.

\subsubsection{$4 \times 4 \times 8$ CuAu alloy out-painting}
$4 \times 4 \times 8$ CuAu lattices were generated using both ARM MLP and MAM Transformer $4 \times 4 \times 4$ CuAu models. We extended the $4 \times 4 \times 4$ lattice along the $z$ dimension by iteratively adding layers with stride 3, following a raster scan order along $z$. Each new $4 \times 4 \times 4$ block was conditionally sampled given the previously filled sites in neighboring blocks, with periodic boundary conditions applied in the $x$ and $y$ dimensions. A single batch of 1000 samples were generated at each condition.

\subsection{Energy models}
\subsubsection{Ising model}
We describe Ising models with $E = -\frac{1}{2}(\bx^\top \mathbf{J} \bx + B \bx)$, where $\bx \in \{-1, 1\}^L$ represents the spin state, $\mathbf{J}$ is a product of interaction strength $\varepsilon = 1$ and the adjacency matrix, and $B$ the field strength.

\subsubsection{CuAu energy model}
We describe CuAu models with $E = E_\text{config} - \mu (N_\text{Au} - N_\text{Cu})$, where $E_\text{config}$ is the configurational energy, $\mu$ is the chemical potential, $N_\text{Au}$ and $N_\text{Cu}$ are the number of Au and Cu atoms, respectively. The energy model was calculated using the CLEASE cluster expansion package~\cite{chang_clease_2019} with parameters from Ref.~\citenum{damewood_sampling_2022}.

\subsection{Observables calculations}
\subsubsection{Ising model two-point correlation function calculations}
We computed the shifted (connected) two-point correlation function to characterize spatial correlations between spins:
\begin{align*}
    C(r) = \langle s_i \cdot s_j \rangle - \langle s_i \rangle \cdot \langle s_j \rangle
\end{align*}
where $s_i$ and $s_j$ are spin variables at sites separated by Manhattan distance $r$, and $\langle \cdot \rangle$ denotes the ensemble average over samples. The subtraction of $\langle s_i \rangle \cdot \langle s_j \rangle$ removes the contribution from mean magnetization, yielding the connected correlation that decays to zero at large distances. Periodic boundary conditions were applied when computing distances. For samples at each thermodynamic condition, we computed $C(r)$ by averaging over all pairs of sites at distance $r$. Calculations were performed for $10 \times 10$ Ising systems at $k_\text{B} T \in [1.5, 3.5]$ in intervals of 0.1 $k_\text{B} T$ and $B \in \{0.0, \pm0.2, \pm0.4\}$.

\subsubsection{Ising model Wang--Landau reference sampling}
The Wang--Landau algorithm~\citep{wang_efficient_2001} was used to obtain density-of-state estimates, $g(E)$, as a function of energy $E$. We performed Wang--Landau sampling at zero field ($B = 0$) for lattice sizes $4 \times 4$, $5 \times 5$, $8 \times 8$, $10 \times 10$, $15 \times 15$, and $20 \times 20$. The algorithm used an initial modification factor $\ln f_{\text{init}} = 1.0$, reduced by half whenever the histogram flatness criterion (minimum bin count $> 0.90 \times$ mean count) was satisfied, checking every 10,000 Monte Carlo steps. Convergence was achieved when $\ln f < 10^{-8}$. 

From the converged density of states at $B=0$, thermodynamic observables at any temperature were computed using canonical ensemble averaging. The partition function at temperature $T$ was calculated as:
\begin{align*}
    Z = \sum_E g(E) \exp(-\beta E)
\end{align*}
where $\beta = 1/(k_\text{B} T)$. The log-sum-exp trick was used for numerical stability in implementation. The free energy per site is: $F/L = -k_\text{B} T \ln Z / L$. The specific heat capacity per site was computed from energy fluctuations:
\begin{align*}
    C_V/L &= \frac{1}{k_\text{B} T^2} \frac{\langle E^2 \rangle - \langle E \rangle^2}{L}
\end{align*}
where thermal averages $\langle E^n \rangle = \sum_E E^n g(E) \exp(-\beta E) / Z$. We calculated for $k_\text{B} T \in [1.5, 3.5]$ in intervals of 0.1 $k_\text{B} T$. Meanwhile, free energy per site for $B\in\{\pm0.2, \pm0.4\}$ were taken from the results of Ref.~\citenum{damewood_sampling_2022} due to high computational cost.

\subsubsection{CuAu alloy exact free energy calculations}
Exact free energies were calculated for the $2 \times 2 \times 4$ CuAu supercell by enumerating all $2^{16} = 65,536$ possible configurations to construct the partition function. The partition function was calculated as:
\begin{align*}
    Z = \sum_{\text{all } \bx}  \exp(-\beta E(\bx))
\end{align*}
where $E$ is the total energy of each configuration. The log-sum-exp trick was used for numerical stability in implementation. The free energy per site is then: $F/L = -k_\text{B} T \ln Z / L$, where $L = 16$ is the number of sites in the supercell.

Exact free energies were computed for $T \in$ [200 K, 1200 K] in intervals of 100 K (11 temperatures) and $\mu \in  \{0.00, \pm0.03, \pm0.06, \pm0.09, \pm0.12\}$ eV (9 values), yielding a total of 99 thermodynamic conditions.

\subsubsection{CuAu reference phase diagrams and free energies with metadynamics sampling}
Metadynamics was performed following the procedure of Ref.~\citenum{damewood_sampling_2022} using CLEASE~\citep{chang_clease_2019}. The collective variable for the metadynamics runs is the concentration of Au, with 65 and 129 bins, one for each possible concentration of Au in the $4 \times 4 \times 4$ and $4 \times 4 \times 8$ systems respectively. Each new sample was proposed using semigrand canonical Monte Carlo and accepted according to the Metropolis-Hastings algorithm. Metadynamics was run at 32 temperatures uniformly spread across $T \in$ [200 K, 1200 K]. At each temperature, each metadynamics step consists of running the sampling for a maximum of 20,000 sweeps until the convergence criterion was met: when the most infrequent bin was visited at least 80\% of the average. The artificial potential ($V$) is initially modified by 10,000 $k_\text{B} T$ when the sampler visits a bin, and is halved at each metadynamics step until $0.0001$ to ensure convergence of the free energy curve. 

Following the metadynamics runs, the points on the convex hull of $V$ across all bins at each temperature were taken to construct the reference temperature-composition phase diagram. The two-phase equilibria boundary was then extracted from the metadynamics phase diagram by linearly connecting the sampled points with the lowest temperature at each Au concentration to obtain the metadynamics reference lines in Fig.~\ref{fig:cuau_phase} and Fig.~S18. Free energies were extracted from the metadynamics potential at the closest concentration bins to target compositions $x_\text{Au} \in \{0.0, 0.25, 0.5, 0.75, 1.0\}$.

\subsubsection{CuAu alloy phase diagrams with ARM/MAM samples}
Phase diagrams were constructed from ARM and MAM model-generated samples by extracting $x_\text{Au}$ for each sample at each thermodynamic condition ($T, \mu$). Temperature-composition phase diagrams were generated by plotting each sample as a point with $x_\text{Au}$ on the horizontal axis and temperature on the vertical axis. Similar to the metadynamics reference phase diagrams, the two-phase equilibria boundaries were extracted from the ARM/MAM phase diagrams by linearly connecting the sampled points with the lowest temperature at each Au concentration using Ref.~\citenum{noauthor_automerisio_nodate}.

\subsubsection{MCMC reference sampling}
MCMC reference samples were generated using Metropolis-Hastings sampling with uniform single-spin flip proposals. For the Ising model, for each thermodynamic condition, we collected samples in batches of 10,000, with 250 MCMC steps between consecutive samples to reduce correlation. The number of batches was increased in increments of 40 until statistical convergence was reached for each lattice size ($10 \times 10$, $15 \times 15$, and $20 \times 20$). Sampling was performed at $k_\text{B} T \in \{1.5, 2.0, 2.5, 3.0, 3.5\}$ and $B \in \{0.0, \pm0.2, \pm0.4\}$ for all lattice sizes. For the $20 \times 20$ Ising model at $k_\text{B} T \in \{1.5, 2.0, 2.5\}$ and $B = 0.0$, we used an alternative sampling strategy with 5,000 MCMC steps between consecutive samples. For the CuAu model, we collected samples in batches of 1,000, with 250 MCMC steps between consecutive samples to reduce correlation. The number of batches was increased in increments of 40 until statistical convergence was reached for all lattice sizes ($2 \times 2 \times 4$, $4 \times 4 \times 4$, and $4 \times 4 \times 8$). Sampling was performed at $T \in$ [200 K, 1200 K] in intervals of 200 K and $\mu \in \{0.00, \pm0.06, \pm0.12\}$ eV (30 unique conditions).

\subsubsection{Free energy estimation with importance sampling}
Using our exact likelihood models, the estimated partition function, $\hat{Z}$, and hence the lattice free energy can be obtained using an asymptotically-unbiased importance sampling method~\citep{nicoli_asymptotically_2020}:
\begin{align*}
    \hat{Z}_N = \frac{1}{N} \sum_{i=1}^N w_i, \quad w_i = \frac{f(\bx_i)}{p(\bx_i)}
\end{align*}
where $\bx_i \sim p(\bx)$ are samples from the learned distribution. The log-sum-exp trick was used for numerical stability in implementation. The free energy is: $\hat{A} = -k_\text{B} T \ln \hat{Z}_N$, normalized by the number of sites. 

For the Ising model at all lattice sizes, we obtained $N = 10,000$ samples for each thermodynamic condition. For $10 \times 10$ Ising systems, we evaluated at $k_\text{B} T$ from 1.5 to 3.5 in intervals of 0.1 $k_{\text{B}}T$ and $B \in \{0.0, \pm0.2, \pm0.4\}$. For $15 \times 15$ and $20 \times 20$ Ising systems, we used $k_\text{B} T \in \{1.5, 2.0, 2.1, 2.2, 2.3, 2.4, 2.5, 2.6, 2.7, 2.8, 2.9, 3.0, 3.5\}$ and $B \in \{0.0, \pm0.2, \pm0.4\}$.

For the CuAu model, we obtained $N = 1,000$ samples for each thermodynamic condition. For all CuAu systems, we evaluated at $T \in$ [200 K, 1200 K] in intervals of 100 K and $\mu \in  \{0.00, \pm0.03, \pm0.06, \pm0.09, \pm0.12\}$ eV (99 unique conditions).

For the $4 \times 4 \times 4$ and $4 \times 4 \times 8$ CuAu systems, free energies at specific compositions were calculated from the sampled configurations. For each sample $i$ at thermodynamic condition $(T, \mu)$, the free energy per site was computed as $F_i = \Omega(T, \mu) + \mu \cdot x_i$, where $\Omega(T, \mu)$ is the $T$--$\mu$ free-energy per site (identical for all samples at the same $(T, \mu)$ condition) and $x_i$ is the Au fraction of sample $i$. Importance weights $w_i = f(\bx_i)/p(\bx_i)$ were first calculated over all samples at the same $(T, \mu)$ condition. Samples were then grouped by temperature and binned by composition using bins of width $\pm 0.02$ centered at target compositions $x_\text{Au} \in \{0.0, 0.25, 0.5, 0.75, 1.0\}$. Within each $(T, x)$ bin, the weights were normalized as $\hat{w}_i = w_i / \sum_{j \in \text{bin}} w_j$, and free energies were averaged as $F(T, x_\text{bin}) = \sum_i \hat{w}_i F_i$ for all samples $i$ in the bin. Log-space operations were used for numerical stability in implementation.

\subsubsection{Heat capacity calculations with reweighted energies}
The heat capacity was computed from the variance of reweighted energies. Using normalized importance sampling weights $\hat{w}_i = w_i / \sum_j w_j$, where $w_i = f(\bx_i)/p(\bx_i)$ are the importance weights, the weighted averages are:
\begin{align*}
    \langle E \rangle_w = \sum_i \hat{w}_i E_i, \quad \langle E^2 \rangle_w = \sum_i \hat{w}_i E_i^2
\end{align*}
giving the heat capacity:
\begin{align*}
    C_V = \frac{\langle E^2 \rangle_w - \langle E \rangle_w^2}{k_\text{B} T^2}
\end{align*}
where energies $E$ are computed at zero field. Log-space operations were used for numerical stability in computing the normalized weights. Heat capacities were calculated at the same thermodynamic conditions as described in the ``Methods: Free energy estimation with importance sampling'' section above.

\subsubsection{Normalized effective sample size calculations}
NESS quantifies the quality of importance sampling and was computed following Ref.~\citenum{nicoli_asymptotically_2020, damewood_sampling_2022}:
\begin{align*}
    \text{NESS} = \frac{(\sum_i w_i)^2}{N \sum_i w_i^2}
\end{align*}
where $w_i = f(\bx_i)/p(\bx_i)$ are the importance weights, and $N$ is the number of samples. Log-space operations were used for numerical stability in implementation. NESS ranges from 0 to 1, with values near 1 indicating high-quality samples. NESS was evaluated at the same thermodynamic conditions as described in the ``Methods: Free energy estimation with importance sampling'' section above.



\section{Data availability}
Data will be made available on Zenodo upon acceptance.

\section{Code availability}
Code will be made available on GitHub upon acceptance.

\begin{acknowledgments}
The authors appreciate manuscript editing by Hoje Chun. X.D. acknowledges support from the National Science Foundation Graduate Research Fellowship under Grant No. 2141064. J.N. acknowledges support from the Mathworks Fellowship. The authors acknowledge the MIT SuperCloud and Lincoln Laboratory Supercomputing Center for providing HPC resources that have contributed to the research results reported within this paper. This research used resources of the National Energy Research Scientific Computing Center (NERSC), a Department of Energy Office of Science User Facility using NERSC awards BES-ERCAP-m4737, DDR-ERCAP-m4866 (AI4Sci@NERSC), and ALCC-ERCAP-m5068.
\end{acknowledgments}

\section*{Supplemental Material}
(1) A list of abbreviations; (2) derivation of the variational free energy objective used for training; and (3) supplementary figures, including representative spin configurations, two-point correlation functions, interaction energy distributions, and up-spin concentration distributions for $10\times10$, $15\times15$, and $20\times20$ Ising systems, normalized effective sample size (NESS) maps across thermodynamic conditions for Ising systems, free energy deviations and additional free energy metrics for CuAu systems, interaction energy and Au concentration distributions for $2\times2\times4$, $4\times4\times4$, and $4\times4\times8$ CuAu systems, temperature--composition phase diagrams for CuAu systems, and NESS maps across thermodynamic conditions for CuAu systems.

\bibliography{references}

@misc{ou_non-arrhenius_2025,
	title = {Non-{Arrhenius} {Li}-ion transport and grain-size effects in argyrodite solid electrolytes},
	url = {http://arxiv.org/abs/2510.18630},
	doi = {10.48550/arXiv.2510.18630},
	urldate = {2026-01-31},
	publisher = {arXiv},
	author = {Ou, Yongliang and Scholz, Lena and Keshav, Sanath and Ikeda, Yuji and Kraft, Marvin and Divinski, Sergiy and Gómez-Bombarelli, Rafael and Zeier, Wolfgang G. and Fritzen, Felix and Grabowski, Blazej},
	month = oct,
	year = {2025},
}

@book{tuckerman2023statistical,
	series = {Oxford graduate texts},
	title = {Statistical mechanics: {Theory} and molecular simulation},
	isbn = {978-0-19-255961-6},
	url = {https://books.google.com/books?id=eLXLEAAAQBAJ},
	publisher = {OUP Oxford},
	author = {Tuckerman, M.E.},
	year = {2023},
}

@book{chandler1987introduction,
	title = {Introduction to modern statistical mechanics},
	isbn = {978-0-19-504277-1},
	url = {https://books.google.com/books?id=3taTh5D-CDsC},
	publisher = {Oxford University Press},
	author = {Chandler, D.},
	year = {1987},
}

@article{van_de_walle_alloy_2002,
	title = {The alloy theoretic automated toolkit: {A} user guide},
	volume = {26},
	issn = {0364-5916},
	shorttitle = {The alloy theoretic automated toolkit},
	url = {https://www.sciencedirect.com/science/article/pii/S0364591602800062},
	doi = {10.1016/S0364-5916(02)80006-2},
	number = {4},
	urldate = {2025-12-23},
	journal = {Calphad},
	author = {van de Walle, A. and Asta, M. and Ceder, G.},
	month = dec,
	year = {2002},
	pages = {539--553},
}

@article{angqvist_icet_2019,
	title = {{ICET} – {A} {Python} {Library} for {Constructing} and {Sampling} {Alloy} {Cluster} {Expansions}},
	volume = {2},
	copyright = {© 2019 WILEY-VCH Verlag GmbH \& Co. KGaA, Weinheim},
	issn = {2513-0390},
	url = {https://onlinelibrary.wiley.com/doi/abs/10.1002/adts.201900015},
	doi = {10.1002/adts.201900015},
	number = {7},
	urldate = {2025-12-23},
	journal = {Advanced Theory and Simulations},
	author = {Ångqvist, Mattias and Muñoz, William A. and Rahm, J. Magnus and Fransson, Erik and Durniak, Céline and Rozyczko, Piotr and Rod, Thomas H. and Erhart, Paul},
	year = {2019},
	pages = {1900015},
}

@article{van_der_ven_first-principles_2018,
	title = {First-{Principles} {Statistical} {Mechanics} of {Multicomponent} {Crystals}},
	volume = {48},
	issn = {1531-7331, 1545-4118},
	url = {https://www.annualreviews.org/doi/10.1146/annurev-matsci-070317-124443},
	doi = {10.1146/annurev-matsci-070317-124443},
	number = {1},
	urldate = {2025-12-23},
	journal = {Annual Review of Materials Research},
	author = {Van Der Ven, A. and Thomas, J.C. and Puchala, B. and Natarajan, A.R.},
	month = jul,
	year = {2018},
	pages = {27--55},
}

@article{xu_reactive_2024,
	title = {Reactive {Processing} of {Furan}-{Based} {Monomers} via {Frontal} {Ring}-{Opening} {Metathesis} {Polymerization} for {High} {Performance} {Materials}},
	volume = {36},
	copyright = {© 2024 The Author(s). Advanced Materials published by Wiley-VCH GmbH},
	issn = {1521-4095},
	url = {https://onlinelibrary.wiley.com/doi/abs/10.1002/adma.202405736},
	doi = {10.1002/adma.202405736},
	number = {36},
	urldate = {2025-12-23},
	journal = {Advanced Materials},
	author = {Xu, Zhenchuang and Chua, Lauren and Singhal, Avni and Krishnan, Pranav and Lessard, Jacob J. and Suslick, Benjamin A. and Chen, Valerie and Sottos, Nancy R. and Gomez-Bombarelli, Rafael and Moore, Jeffrey S.},
	year = {2024},
	pages = {2405736},
}

@article{gartner_modeling_2019,
	title = {Modeling and {Simulations} of {Polymers}: {A} {Roadmap}},
	volume = {52},
	issn = {0024-9297},
	shorttitle = {Modeling and {Simulations} of {Polymers}},
	url = {https://doi.org/10.1021/acs.macromol.8b01836},
	doi = {10.1021/acs.macromol.8b01836},
	number = {3},
	urldate = {2025-12-23},
	journal = {Macromolecules},
	publisher = {American Chemical Society},
	author = {Gartner, Thomas E. III and Jayaraman, Arthi},
	month = feb,
	year = {2019},
	pages = {755--786},
}

@article{ou_atomistic_2024,
	title = {Atomistic modeling of bulk and grain boundary diffusion in solid electrolyte $\mathrm{Li}_{6}\mathrm{PS}_{5}\mathrm{Cl}$ using machine-learning interatomic potentials},
	volume = {8},
	url = {https://link.aps.org/doi/10.1103/PhysRevMaterials.8.115407},
	doi = {10.1103/PhysRevMaterials.8.115407},
	number = {11},
	urldate = {2025-12-23},
	journal = {Physical Review Materials},
	publisher = {American Physical Society},
	author = {Ou, Yongliang and Ikeda, Yuji and Scholz, Lena and Divinski, Sergiy and Fritzen, Felix and Grabowski, Blazej},
	month = nov,
	year = {2024},
	pages = {115407},
}

@article{sheriff_chemical-motif_2024,
	title = {Chemical-motif characterization of short-range order with {E}(3)-equivariant graph neural networks},
	volume = {10},
	copyright = {2024 The Author(s)},
	issn = {2057-3960},
	url = {https://www.nature.com/articles/s41524-024-01393-5},
	doi = {10.1038/s41524-024-01393-5},
	number = {1},
	urldate = {2025-12-23},
	journal = {npj Computational Materials},
	publisher = {Nature Publishing Group},
	author = {Sheriff, Killian and Cao, Yifan and Freitas, Rodrigo},
	month = sep,
	year = {2024},
	pages = {215},
}

@article{yang_learning_2024,
	title = {Learning {Collective} {Variables} with {Synthetic} {Data} {Augmentation} through {Physics}-{Inspired} {Geodesic} {Interpolation}},
	volume = {20},
	issn = {1549-9618},
	url = {https://doi.org/10.1021/acs.jctc.4c00435},
	doi = {10.1021/acs.jctc.4c00435},
	number = {15},
	urldate = {2025-12-23},
	journal = {Journal of Chemical Theory and Computation},
	publisher = {American Chemical Society},
	author = {Yang, Soojung and Nam, Juno and Dietschreit, Johannes C. B. and Gómez-Bombarelli, Rafael},
	month = aug,
	year = {2024},
	pages = {6559--6568},
}

@misc{noauthor_automerisio_nodate,
	title = {automeris.io: {Computer} vision assisted data extraction from charts using {WebPlotDigitizer}},
	url = {https://automeris.io/},
	urldate = {2025-12-22},
}

@article{herron_inferring_2024,
	title = {Inferring phase transitions and critical exponents from limited observations with thermodynamic maps},
	volume = {121},
	url = {https://www.pnas.org/doi/10.1073/pnas.2321971121},
	doi = {10.1073/pnas.2321971121},
	number = {52},
	urldate = {2025-12-04},
	journal = {Proceedings of the National Academy of Sciences},
	publisher = {Proceedings of the National Academy of Sciences},
	author = {Herron, Lukas and Mondal, Kinjal and Schneekloth, John S. and Tiwary, Pratyush},
	month = dec,
	year = {2024},
	pages = {e2321971121},
}

@article{tuo_scalable_2025,
	title = {Scalable {Multitemperature} {Free} {Energy} {Sampling} of {Classical} {Ising} {Spin} {States}},
	volume = {21},
	issn = {1549-9618},
	url = {https://doi.org/10.1021/acs.jctc.5c01248},
	doi = {10.1021/acs.jctc.5c01248},
	number = {22},
	urldate = {2025-12-04},
	journal = {Journal of Chemical Theory and Computation},
	publisher = {American Chemical Society},
	author = {Tuo, Ping and Zeng, Zezhu and Chen, Jiale and Cheng, Bingqing},
	month = nov,
	year = {2025},
	pages = {11427--11435},
}

@misc{vaswani_attention_2023,
	title = {Attention {Is} {All} {You} {Need}},
	url = {http://arxiv.org/abs/1706.03762},
	doi = {10.48550/arXiv.1706.03762},
	urldate = {2025-12-04},
	publisher = {arXiv},
	author = {Vaswani, Ashish and Shazeer, Noam and Parmar, Niki and Uszkoreit, Jakob and Jones, Llion and Gomez, Aidan N. and Kaiser, Lukasz and Polosukhin, Illia},
	month = aug,
	year = {2023},
}

@article{nam_transferable_2025,
	title = {Transferable {Learning} of {Reaction} {Pathways} from {Geometric} {Priors}},
	volume = {16},
	url = {https://doi.org/10.1021/acs.jpclett.5c02620},
	doi = {10.1021/acs.jpclett.5c02620},
	number = {45},
	urldate = {2025-11-19},
	journal = {The Journal of Physical Chemistry Letters},
	publisher = {American Chemical Society},
	author = {Nam, Juno and Steiner, Miguel and Misterka, Max and Yang, Soojung and Singhal, Avni and Gómez-Bombarelli, Rafael},
	month = nov,
	year = {2025},
	pages = {11690--11699},
}

@misc{guo_proximal_2025,
	title = {Proximal {Diffusion} {Neural} {Sampler}},
	url = {http://arxiv.org/abs/2510.03824},
	doi = {10.48550/arXiv.2510.03824},
	urldate = {2025-10-31},
	publisher = {arXiv},
	author = {Guo, Wei and Choi, Jaemoo and Zhu, Yuchen and Tao, Molei and Chen, Yongxin},
	month = oct,
	year = {2025},
}

@article{nam_flow_2025,
	title = {Flow matching for accelerated simulation of atomic transport in crystalline materials},
	copyright = {2025 The Author(s), under exclusive licence to Springer Nature Limited},
	issn = {2522-5839},
	url = {https://www.nature.com/articles/s42256-025-01125-4},
	doi = {10.1038/s42256-025-01125-4},
	urldate = {2025-10-22},
	journal = {Nature Machine Intelligence},
	publisher = {Nature Publishing Group},
	author = {Nam, Juno and Liu, Sulin and Winter, Gavin and Jun, KyuJung and Yang, Soojung and Gómez-Bombarelli, Rafael},
	month = oct,
	year = {2025},
	pages = {1--11},
}

@misc{woo_energy-based_2025,
	title = {Energy-based generator matching: {A} neural sampler for general state space},
	shorttitle = {Energy-based generator matching},
	url = {http://arxiv.org/abs/2505.19646},
	doi = {10.48550/arXiv.2505.19646},
	urldate = {2025-10-10},
	publisher = {arXiv},
	author = {Woo, Dongyeop and Kim, Minsu and Kim, Minkyu and Seong, Kiyoung and Ahn, Sungsoo},
	month = may,
	year = {2025},
}

@misc{ou_discrete_2025,
	title = {Discrete {Neural} {Flow} {Samplers} with {Locally} {Equivariant} {Transformer}},
	url = {http://arxiv.org/abs/2505.17741},
	doi = {10.48550/arXiv.2505.17741},
	urldate = {2025-10-09},
	publisher = {arXiv},
	author = {Ou, Zijing and Zhang, Ruixiang and Li, Yingzhen},
	month = may,
	year = {2025},
}

@misc{zhu_mdns_2025,
	title = {{MDNS}: {Masked} {Diffusion} {Neural} {Sampler} via {Stochastic} {Optimal} {Control}},
	shorttitle = {{MDNS}},
	url = {http://arxiv.org/abs/2508.10684},
	doi = {10.48550/arXiv.2508.10684},
	urldate = {2025-10-09},
	publisher = {arXiv},
	author = {Zhu, Yuchen and Guo, Wei and Choi, Jaemoo and Liu, Guan-Horng and Chen, Yongxin and Tao, Molei},
	month = aug,
	year = {2025},
}

@article{du_accelerating_2025,
	title = {Accelerating and {Enhancing} {Thermodynamic} {Simulations} of {Electrochemical} {Interfaces}},
	volume = {11},
	copyright = {https://creativecommons.org/licenses/by/4.0/},
	issn = {2374-7943, 2374-7951},
	url = {https://pubs.acs.org/doi/10.1021/acscentsci.5c00547},
	doi = {10.1021/acscentsci.5c00547},
	number = {9},
	urldate = {2025-10-02},
	journal = {ACS Central Science},
	author = {Du, Xiaochen and Liu, Mengren and Peng, Jiayu and Chun, Hoje and Hoffman, Alexander and Yildiz, Bilge and Li, Lin and Bazant, Martin Z. and Gómez-Bombarelli, Rafael},
	month = sep,
	year = {2025},
	pages = {1558--1572},
}

@article{du_machine-learning-accelerated_2023,
	title = {Machine-learning-accelerated simulations to enable automatic surface reconstruction},
	copyright = {2023 The Author(s), under exclusive licence to Springer Nature America, Inc.},
	issn = {2662-8457},
	url = {https://www.nature.com/articles/s43588-023-00571-7},
	doi = {10.1038/s43588-023-00571-7},
	urldate = {2023-12-07},
	journal = {Nature Computational Science},
	publisher = {Nature Publishing Group},
	author = {Du, Xiaochen and Damewood, James K. and Lunger, Jaclyn R. and Millan, Reisel and Yildiz, Bilge and Li, Lin and Gómez-Bombarelli, Rafael},
	month = dec,
	year = {2023},
	pages = {1034--1044},
}

@misc{sheriff_machine_2025,
	title = {Machine learning potentials for modeling alloys across compositions},
	url = {http://arxiv.org/abs/2506.12592},
	doi = {10.48550/arXiv.2506.12592},
	urldate = {2025-06-17},
	publisher = {arXiv},
	author = {Sheriff, Killian and Xiao, Daniel and Cao, Yifan and Owen, Lewis R. and Freitas, Rodrigo},
	month = jun,
	year = {2025},
}

@article{spotte-smith_toward_2022,
	title = {Toward a {Mechanistic} {Model} of {Solid}–{Electrolyte} {Interphase} {Formation} and {Evolution} in {Lithium}-{Ion} {Batteries}},
	volume = {7},
	url = {https://doi.org/10.1021/acsenergylett.2c00517},
	doi = {10.1021/acsenergylett.2c00517},
	number = {4},
	urldate = {2025-06-03},
	journal = {ACS Energy Letters},
	publisher = {American Chemical Society},
	author = {Spotte-Smith, Evan Walter Clark and Kam, Ronald L. and Barter, Daniel and Xie, Xiaowei and Hou, Tingzheng and Dwaraknath, Shyam and Blau, Samuel M. and Persson, Kristin A.},
	month = apr,
	year = {2022},
	pages = {1446--1453},
}

@article{lee_machine-learning-accelerated_2025,
	title = {Machine-{Learning}-{Accelerated} {Surface} {Exploration} of {Reconstructed} {BiVO4}(010) and {Characterization} of {Their} {Aqueous} {Interfaces}},
	volume = {147},
	issn = {0002-7863},
	url = {https://doi.org/10.1021/jacs.4c17739},
	doi = {10.1021/jacs.4c17739},
	number = {9},
	urldate = {2025-05-17},
	journal = {Journal of the American Chemical Society},
	publisher = {American Chemical Society},
	author = {Lee, Yonghyuk and Lee, Taehun},
	month = mar,
	year = {2025},
	pages = {7799--7808},
}

@misc{chun_learning_2024,
	title = {Learning {Mean} {First} {Passage} {Time}: {Chemical} {Short}-{Range} {Order} and {Kinetics} of {Diffusive} {Relaxation}},
	shorttitle = {Learning {Mean} {First} {Passage} {Time}},
	url = {http://arxiv.org/abs/2411.17839},
	doi = {10.48550/arXiv.2411.17839},
	urldate = {2025-05-16},
	publisher = {arXiv},
	author = {Chun, Hoje and Tang, Hao and Gomez-Bombarelli, Rafael and Li, Ju},
	month = nov,
	year = {2024},
}

@misc{lee_exponentially_2025,
	title = {Exponentially {Tilted} {Thermodynamic} {Maps} ({expTM}): {Predicting} {Phase} {Transitions} {Across} {Temperature}, {Pressure}, and {Chemical} {Potential}},
	shorttitle = {Exponentially {Tilted} {Thermodynamic} {Maps} ({expTM})},
	url = {http://arxiv.org/abs/2503.15080},
	doi = {10.48550/arXiv.2503.15080},
	urldate = {2025-04-01},
	publisher = {arXiv},
	author = {Lee, Suemin and Wang, Ruiyu and Herron, Lukas and Tiwary, Pratyush},
	month = mar,
	year = {2025},
}

@article{peng_data-driven_2024,
	title = {Data-driven physics-informed descriptors of cation ordering in multicomponent perovskite oxides},
	volume = {5},
	issn = {2666-3864},
	url = {https://www.cell.com/cell-reports-physical-science/abstract/S2666-3864(24)00191-7},
	doi = {10.1016/j.xcrp.2024.101942},
	number = {5},
	urldate = {2025-03-13},
	journal = {Cell Reports Physical Science},
	publisher = {Elsevier},
	author = {Peng, Jiayu and Damewood, James and Gómez-Bombarelli, Rafael},
	month = may,
	year = {2024},
}

@misc{holderrieth_leaps_2025,
	title = {{LEAPS}: {A} discrete neural sampler via locally equivariant networks},
	shorttitle = {{LEAPS}},
	url = {http://arxiv.org/abs/2502.10843},
	doi = {10.48550/arXiv.2502.10843},
	urldate = {2025-02-24},
	publisher = {arXiv},
	author = {Holderrieth, Peter and Albergo, Michael S. and Jaakkola, Tommi},
	month = feb,
	year = {2025},
}

@article{peng_stability_2022,
	title = {Stability {Design} {Principles} of {Manganese}-{Based} {Oxides} in {Acid}},
	volume = {34},
	issn = {0897-4756},
	url = {https://doi.org/10.1021/acs.chemmater.2c01233},
	doi = {10.1021/acs.chemmater.2c01233},
	number = {17},
	urldate = {2025-02-10},
	journal = {Chemistry of Materials},
	publisher = {American Chemical Society},
	author = {Peng, Jiayu and Giordano, Livia and Davenport, Timothy C. and Shao-Horn, Yang},
	month = sep,
	year = {2022},
	pages = {7774--7787},
}

@article{lunger_towards_2024,
	title = {Towards atom-level understanding of metal oxide catalysts for the oxygen evolution reaction with machine learning},
	volume = {10},
	copyright = {2024 The Author(s)},
	issn = {2057-3960},
	url = {https://www.nature.com/articles/s41524-024-01273-y},
	doi = {10.1038/s41524-024-01273-y},
	number = {1},
	urldate = {2025-01-31},
	journal = {npj Computational Materials},
	publisher = {Nature Publishing Group},
	author = {Lunger, Jaclyn R. and Karaguesian, Jessica and Chun, Hoje and Peng, Jiayu and Tseo, Yitong and Shan, Chung Hsuan and Han, Byungchan and Shao-Horn, Yang and Gómez-Bombarelli, Rafael},
	month = apr,
	year = {2024},
	pages = {1--11},
}

@article{kempen_inverse_2025,
	title = {Inverse catalysts: tuning the composition and structure of oxide clusters through the metal support},
	volume = {11},
	copyright = {2025 The Author(s)},
	issn = {2057-3960},
	shorttitle = {Inverse catalysts},
	url = {https://www.nature.com/articles/s41524-024-01507-z},
	doi = {10.1038/s41524-024-01507-z},
	number = {1},
	urldate = {2025-01-20},
	journal = {npj Computational Materials},
	publisher = {Nature Publishing Group},
	author = {Kempen, Luuk H. E. and Andersen, Mie},
	month = jan,
	year = {2025},
	pages = {1--12},
}

@article{chang_clease_2019,
	title = {{CLEASE}: a versatile and user-friendly implementation of cluster expansion method},
	volume = {31},
	issn = {0953-8984},
	shorttitle = {{CLEASE}},
	url = {https://dx.doi.org/10.1088/1361-648X/ab1bbc},
	doi = {10.1088/1361-648X/ab1bbc},
	number = {32},
	urldate = {2024-10-18},
	journal = {Journal of Physics: Condensed Matter},
	publisher = {IOP Publishing},
	author = {Chang, Jin Hyun and Kleiven, David and Melander, Marko and Akola, Jaakko and Garcia-Lastra, Juan Maria and Vegge, Tejs},
	month = may,
	year = {2019},
	pages = {325901},
}

@misc{chun_active_2024,
	title = {Active learning accelerated exploration of single-atom local environments in multimetallic systems for oxygen electrocatalysis},
	url = {https://chemrxiv.org/engage/chemrxiv/article-details/65f09f1b9138d2316153e3f4},
	doi = {10.26434/chemrxiv-2024-6skxn},
	urldate = {2024-10-13},
	publisher = {ChemRxiv},
	author = {Chun, Hoje and Lunger, Jaclyn and Kang, Jeung Ku and Gómez-Bombarelli, Rafael and Han, Byungchan},
	month = mar,
	year = {2024},
}

@article{li_origin_2022,
	title = {Origin of the herringbone reconstruction of {Au}(111) surface at the atomic scale},
	volume = {8},
	url = {https://www.science.org/doi/10.1126/sciadv.abq2900},
	doi = {10.1126/sciadv.abq2900},
	number = {40},
	urldate = {2024-10-13},
	journal = {Science Advances},
	publisher = {American Association for the Advancement of Science},
	author = {Li, Pai and Ding, Feng},
	month = oct,
	year = {2022},
	pages = {eabq2900},
}

@article{han_multifunctional_2024,
	title = {Multifunctional high-entropy materials},
	copyright = {2024 Springer Nature Limited},
	issn = {2058-8437},
	url = {https://www.nature.com/articles/s41578-024-00720-y},
	doi = {10.1038/s41578-024-00720-y},
	urldate = {2024-10-05},
	journal = {Nature Reviews Materials},
	publisher = {Nature Publishing Group},
	author = {Han, Liuliu and Zhu, Shuya and Rao, Ziyuan and Scheu, Christina and Ponge, Dirk and Ludwig, Alfred and Zhang, Hongbin and Gutfleisch, Oliver and Hahn, Horst and Li, Zhiming and Raabe, Dierk},
	month = sep,
	year = {2024},
	pages = {1--20},
}

@article{owen_low-index_2024,
	title = {Low-index mesoscopic surface reconstructions of {Au} surfaces using {Bayesian} force fields},
	volume = {15},
	copyright = {2024 The Author(s)},
	issn = {2041-1723},
	url = {https://www.nature.com/articles/s41467-024-48192-6},
	doi = {10.1038/s41467-024-48192-6},
	number = {1},
	urldate = {2024-09-26},
	journal = {Nature Communications},
	publisher = {Nature Publishing Group},
	author = {Owen, Cameron J. and Xie, Yu and Johansson, Anders and Sun, Lixin and Kozinsky, Boris},
	month = may,
	year = {2024},
	pages = {3790},
}

@article{sheriff_quantifying_2024,
	title = {Quantifying chemical short-range order in metallic alloys},
	volume = {121},
	url = {https://www.pnas.org/doi/abs/10.1073/pnas.2322962121},
	doi = {10.1073/pnas.2322962121},
	number = {25},
	urldate = {2024-09-03},
	journal = {Proceedings of the National Academy of Sciences},
	publisher = {Proceedings of the National Academy of Sciences},
	author = {Sheriff, Killian and Cao, Yifan and Smidt, Tess and Freitas, Rodrigo},
	month = jun,
	year = {2024},
	pages = {e2322962121},
}

@misc{su_roformer_2023,
	title = {{RoFormer}: {Enhanced} {Transformer} with {Rotary} {Position} {Embedding}},
	shorttitle = {{RoFormer}},
	url = {http://arxiv.org/abs/2104.09864},
	doi = {10.48550/arXiv.2104.09864},
	urldate = {2024-09-02},
	publisher = {arXiv},
	author = {Su, Jianlin and Lu, Yu and Pan, Shengfeng and Murtadha, Ahmed and Wen, Bo and Liu, Yunfeng},
	month = nov,
	year = {2023},
}

@inproceedings{liu_generative_2024,
	title = {Generative {Marginalization} {Models}},
	issn = {2640-3498},
	url = {https://proceedings.mlr.press/v235/liu24az.html},
	urldate = {2024-08-09},
	booktitle = {Proceedings of the 41st {International} {Conference} on {Machine} {Learning}},
	publisher = {PMLR},
	author = {Liu, Sulin and Ramadge, Peter and Adams, Ryan P.},
	month = jul,
	year = {2024},
	pages = {31773--31807},
}

@article{xiao_understanding_2020,
	title = {Understanding interface stability in solid-state batteries},
	volume = {5},
	copyright = {2019 Springer Nature Limited},
	issn = {2058-8437},
	url = {https://www.nature.com/articles/s41578-019-0157-5},
	doi = {10.1038/s41578-019-0157-5},
	number = {2},
	urldate = {2024-02-23},
	journal = {Nature Reviews Materials},
	publisher = {Nature Publishing Group},
	author = {Xiao, Yihan and Wang, Yan and Bo, Shou-Hang and Kim, Jae Chul and Miara, Lincoln J. and Ceder, Gerbrand},
	month = feb,
	year = {2020},
	pages = {105--126},
}

@article{williams_simple_1992,
	title = {Simple statistical gradient-following algorithms for connectionist reinforcement learning},
	volume = {8},
	issn = {1573-0565},
	url = {https://doi.org/10.1007/BF00992696},
	doi = {10.1007/BF00992696},
	number = {3},
	urldate = {2023-12-08},
	journal = {Machine Learning},
	author = {Williams, Ronald J.},
	month = may,
	year = {1992},
	pages = {229--256},
}

@article{seh_combining_2017,
	title = {Combining theory and experiment in electrocatalysis: {Insights} into materials design},
	volume = {355},
	shorttitle = {Combining theory and experiment in electrocatalysis},
	url = {https://www.science.org/doi/10.1126/science.aad4998},
	doi = {10.1126/science.aad4998},
	number = {6321},
	urldate = {2023-12-03},
	journal = {Science},
	publisher = {American Association for the Advancement of Science},
	author = {Seh, Zhi Wei and Kibsgaard, Jakob and Dickens, Colin F. and Chorkendorff, Ib and Nørskov, Jens K. and Jaramillo, Thomas F.},
	month = jan,
	year = {2017},
	pages = {eaad4998},
}

@article{bruix_first-principles-based_2019,
	title = {First-principles-based multiscale modelling of heterogeneous catalysis},
	volume = {2},
	copyright = {2019 Springer Nature Limited},
	issn = {2520-1158},
	url = {https://www.nature.com/articles/s41929-019-0298-3},
	doi = {10.1038/s41929-019-0298-3},
	number = {8},
	urldate = {2023-11-30},
	journal = {Nature Catalysis},
	publisher = {Nature Publishing Group},
	author = {Bruix, Albert and Margraf, Johannes T. and Andersen, Mie and Reuter, Karsten},
	month = aug,
	year = {2019},
	pages = {659--670},
}

@article{takeuchi_new_2017,
	title = {New {Wang}-{Landau} approach to obtain phase diagrams for multicomponent alloys},
	volume = {96},
	url = {https://link.aps.org/doi/10.1103/PhysRevB.96.144202},
	doi = {10.1103/PhysRevB.96.144202},
	number = {14},
	urldate = {2023-11-16},
	journal = {Physical Review B},
	publisher = {American Physical Society},
	author = {Takeuchi, Kazuhito and Tanaka, Ryohei and Yuge, Koretaka},
	month = oct,
	year = {2017},
	pages = {144202},
}

@inproceedings{schutt_equivariant_2021,
	address = {Virtual},
	series = {Proceedings of {Machine} {Learning} {Research}},
	title = {Equivariant message passing for the prediction of tensorial properties and molecular spectra},
	volume = {139},
	url = {https://proceedings.mlr.press/v139/schutt21a.html},
	booktitle = {Proceedings of the 38th international conference on machine learning},
	publisher = {PMLR},
	author = {Schütt, Kristof and Unke, Oliver and Gastegger, Michael},
	editor = {Meila, Marina and Zhang, Tong},
	month = jul,
	year = {2021},
	pages = {9377--9388},
}

@article{nicoli_asymptotically_2020,
	title = {Asymptotically unbiased estimation of physical observables with neural samplers},
	volume = {101},
	url = {https://link.aps.org/doi/10.1103/PhysRevE.101.023304},
	doi = {10.1103/PhysRevE.101.023304},
	number = {2},
	urldate = {2023-10-31},
	journal = {Physical Review E},
	publisher = {American Physical Society},
	author = {Nicoli, Kim A. and Nakajima, Shinichi and Strodthoff, Nils and Samek, Wojciech and Müller, Klaus-Robert and Kessel, Pan},
	month = feb,
	year = {2020},
	pages = {023304},
}

@article{wu_solving_2019,
	title = {Solving {Statistical} {Mechanics} {Using} {Variational} {Autoregressive} {Networks}},
	volume = {122},
	url = {https://link.aps.org/doi/10.1103/PhysRevLett.122.080602},
	doi = {10.1103/PhysRevLett.122.080602},
	number = {8},
	urldate = {2023-10-30},
	journal = {Physical Review Letters},
	publisher = {American Physical Society},
	author = {Wu, Dian and Wang, Lei and Zhang, Pan},
	month = feb,
	year = {2019},
	pages = {080602},
}

@article{niu_multi-cell_2019,
	title = {Multi-cell {Monte} {Carlo} method for phase prediction},
	volume = {5},
	copyright = {2019 The Author(s)},
	issn = {2057-3960},
	url = {https://www.nature.com/articles/s41524-019-0259-z},
	doi = {10.1038/s41524-019-0259-z},
	number = {1},
	urldate = {2023-10-30},
	journal = {npj Computational Materials},
	publisher = {Nature Publishing Group},
	author = {Niu, Changning and Rao, You and Windl, Wolfgang and Ghazisaeidi, Maryam},
	month = dec,
	year = {2019},
	pages = {1--5},
}

@article{wang_efficient_2001,
	title = {Efficient, {Multiple}-{Range} {Random} {Walk} {Algorithm} to {Calculate} the {Density} of {States}},
	volume = {86},
	url = {https://link.aps.org/doi/10.1103/PhysRevLett.86.2050},
	doi = {10.1103/PhysRevLett.86.2050},
	number = {10},
	urldate = {2023-10-30},
	journal = {Physical Review Letters},
	publisher = {American Physical Society},
	author = {Wang, Fugao and Landau, D. P.},
	month = mar,
	year = {2001},
	pages = {2050--2053},
}

@article{shen_deciphering_2023,
	title = {Deciphering the {Atomistic} {Mechanism} of {Si}(111)-7 × 7 {Surface} {Reconstruction} {Using} a {Machine}-{Learning} {Force} {Field}},
	volume = {145},
	issn = {0002-7863},
	url = {https://doi.org/10.1021/jacs.3c06540},
	doi = {10.1021/jacs.3c06540},
	number = {37},
	urldate = {2023-10-05},
	journal = {Journal of the American Chemical Society},
	publisher = {American Chemical Society},
	author = {Shen, Yidi and Morozov, Sergey I. and Luo, Kun and An, Qi and Goddard III, William A.},
	month = sep,
	year = {2023},
	pages = {20511--20520},
}

@article{damewood_sampling_2022,
	title = {Sampling lattices in semi-grand canonical ensemble with autoregressive machine learning},
	volume = {8},
	copyright = {2022 The Author(s)},
	issn = {2057-3960},
	url = {https://www.nature.com/articles/s41524-022-00736-4},
	doi = {10.1038/s41524-022-00736-4},
	number = {1},
	urldate = {2023-03-18},
	journal = {npj Computational Materials},
	publisher = {Nature Publishing Group},
	author = {Damewood, James and Schwalbe-Koda, Daniel and Gómez-Bombarelli, Rafael},
	month = apr,
	year = {2022},
	pages = {1--10},
}

@article{wexler_automatic_2019,
	title = {Automatic {Prediction} of {Surface} {Phase} {Diagrams} {Using} {Ab} {Initio} {Grand} {Canonical} {Monte} {Carlo}},
	volume = {123},
	issn = {1932-7447},
	url = {https://doi.org/10.1021/acs.jpcc.8b11093},
	doi = {10.1021/acs.jpcc.8b11093},
	number = {4},
	urldate = {2023-03-04},
	journal = {The Journal of Physical Chemistry C},
	publisher = {American Chemical Society},
	author = {Wexler, Robert B. and Qiu, Tian and Rappe, Andrew M.},
	month = jan,
	year = {2019},
	pages = {2321--2328},
}

@article{xu_atomistic_2022,
	title = {Atomistic {Insights} into the {Oxidation} of {Flat} and {Stepped} {Platinum} {Surfaces} {Using} {Large}-{Scale} {Machine} {Learning} {Potential}-{Based} {Grand}-{Canonical} {Monte} {Carlo}},
	volume = {12},
	url = {https://doi.org/10.1021/acscatal.2c03976},
	doi = {10.1021/acscatal.2c03976},
	number = {24},
	urldate = {2023-03-04},
	journal = {ACS Catalysis},
	publisher = {American Chemical Society},
	author = {Xu, Jiayan and Xie, Wenbo and Han, Yulan and Hu, P.},
	month = dec,
	year = {2022},
	pages = {14812--14824},
}

\end{document}